\documentclass[%
twocolumn, 
english,
amsmath,amssymb,
aps,
pra,
 floatfix,
]{revtex4-1}

\usepackage{graphicx}
\usepackage{dcolumn}
\usepackage{bm}
\usepackage{physics}

\usepackage[usenames, dvipsnames]{color}
\usepackage{sidecap}
\usepackage{textcomp}

\begin{document}

\title{Fault-tolerant operation of a logical qubit in a diamond quantum processor}

\author{M. H. Abobeih$^{1,2}$}
\author{Y. Wang$^{1}$}
\author{J. Randall$^{1,2}$}
\author{S. J. H. Loenen$^{1,2}$}
\author{C. E. Bradley$^{1,2}$}
\author{M. Markham$^3$} 
\author{D. J. Twitchen$^3$}
\author{B. M. Terhal$^{1,4}$}
\author{T. H. Taminiau$^{1,2}$}
\email{T.H.Taminiau@TUDelft.nl}

\affiliation{$^{1}$QuTech, Delft University of Technology, PO Box 5046, 2600 GA Delft, The Netherlands}
\affiliation{$^{2}$Kavli Institute of Nanoscience Delft, Delft University of Technology, PO Box 5046, 2600 GA Delft, The Netherlands}
\affiliation{$^{3}$Element Six, Fermi Avenue, Harwell Oxford, Didcot, Oxfordshire, OX11 0QR, United Kingdom}
\affiliation{$^4$JARA Institute for Quantum Information, Forschungszentrum Juelich, D-52425 Juelich, Germany}

\date{\today}


\begin{abstract}
Solid-state spin qubits are a promising platform for quantum computation and quantum networks \cite{awschalom2018quantum,chatterjee2021semiconductor}. Recent experiments have demonstrated high-quality control over multi-qubit systems \cite{bradley2019ten,nguyen2019quantum,bourassa2020entanglement,mkadzik2021precision,he2019two,xue2021computing}, elementary quantum algorithms \cite{van2012decoherence,van2019multipartite,vorobyov2020quantum,xue2021computing} and non-fault-tolerant error correction \cite{unden_quantum_2016,waldherr_quantum_2014,cramer_repeated_2016}. Large-scale systems will require using error-corrected logical qubits that are operated fault-tolerantly, so that reliable computation is possible despite noisy operations \cite{preskill_reliable_1998,gottesman_stabilizer_1997,aliferis2005quantum,terhal_quantum_2015}. Overcoming imperfections in this way remains a major outstanding challenge for quantum science \cite{preskill_reliable_1998,nickerson_topological_2013,nigg_quantum_2014,egan2020fault,linke2017fault,rosenblum2018fault,campagne2020quantum,andersen2020repeated,marques2021logical,chen2021exponential}. Here, we demonstrate fault-tolerant operations on a logical qubit using spin qubits in diamond. Our approach is based on the 5-qubit code with a recently discovered flag protocol that enables fault-tolerance using a total of seven qubits \cite{chao2018,chamberland2018flag,chao2020PRXQ}. We encode the logical qubit using a novel protocol based on repeated multi-qubit measurements and show that it outperforms non-fault-tolerant encoding schemes. We then fault-tolerantly manipulate the logical qubit through a complete set of single-qubit Clifford gates. Finally, we demonstrate flagged stabilizer measurements with real-time processing of the outcomes. Such measurements are a primitive for fault-tolerant quantum error correction. While future improvements in fidelity and the number of qubits will be required, our realization of fault-tolerant protocols on the logical-qubit level is a key step towards large-scale quantum information processing based on solid-state spins.
\end{abstract}

\maketitle


\begin{figure*}
\centering
\includegraphics[width=0.9\textwidth]{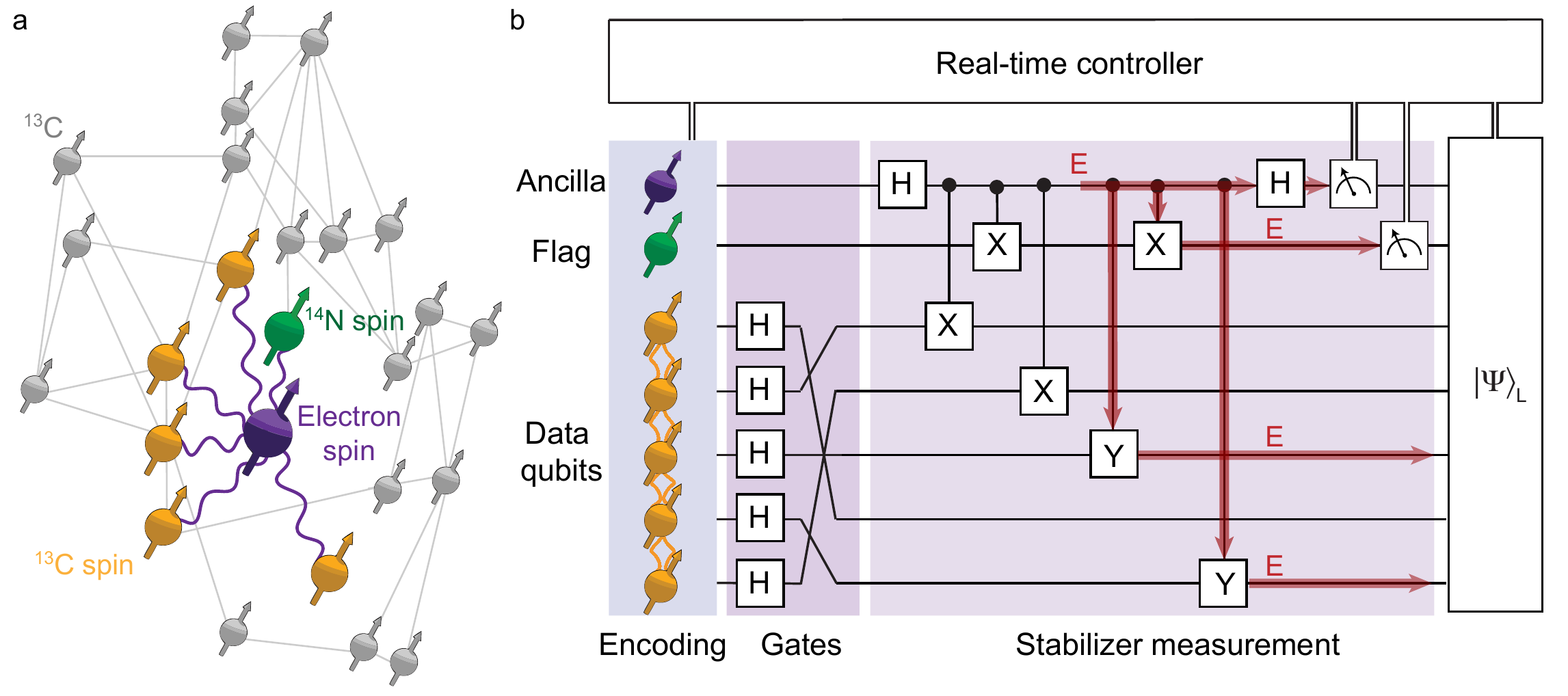}
\caption{Diamond quantum processor, logical qubit and fault-tolerance. a) Our processor consists of a single NV centre and 27 $^{13}$C nuclear-spin qubits, for which the lattice sites and qubit-qubit interactions are known \cite{abobeih2019atomic}. We select 5 $^{13}$C qubits as data qubits that encode the logical state (yellow). The other qubits (grey) are not used here. We use the NV electron spin (purple) as an ancilla qubit for stabilizer measurements, and the NV $^{14}$N nuclear spin (green) as a flag qubit to ensure fault-tolerance. Purple lines: electron-nuclear two-qubit gates used here (Methods). Grey lines: dipolar nuclear-nuclear couplings greater than 6 Hz. b) Illustration of the main components of the experiment. We realize fault-tolerant encoding, gates and stabilizer measurements with real-time processing on a logical qubit of the 5-qubit quantum error correction code. To ensure that any single fault does not cause a logical error, an additional flag qubit is used to identify errors that would propagate to multi-qubit errors and corrupt the logical state \cite{chao2018}. An illustration of such an error $E$ is shown in red.}
\label{fig1}
\end{figure*}


\noindent Large-scale quantum computers and quantum networks will require quantum error correction to overcome inevitable imperfections \cite{gottesman_stabilizer_1997,preskill_reliable_1998,aliferis2005quantum,terhal_quantum_2015,nickerson_topological_2013}. The central idea is to encode each logical qubit of information into multiple physical data qubits. Non-destructive multi-qubit measurements, called stabilizer measurements, can then be used to identify and correct errors \cite{gottesman_stabilizer_1997,preskill_reliable_1998,aliferis2005quantum,terhal_quantum_2015}. If the error rates of all the components are below a certain threshold, it becomes possible to perform arbitrarily large quantum computations by encoding into increasingly more physical qubits \cite{preskill_reliable_1998,aliferis2005quantum,terhal_quantum_2015}. A crucial requirement is that all logical building blocks, including the error-syndrome measurement, must be implemented fault tolerantly. At the lowest level, this implies that any single physical error should not cause a logical error.

Over the last years, steps towards suppressing errors through fault-tolerant quantum error correction have been made in various hardware platforms such as superconducting qubits \cite{rosenblum2018fault,campagne2020quantum,andersen2020repeated,marques2021logical,chen2021exponential}, trapped-ion qubits \cite{nigg_quantum_2014,negnevitsky2018repeated,egan2020fault}, spin qubits in silicon \cite{xue2021computing,he2019two,mkadzik2021precision}, and spin qubits in diamond \cite{waldherr_quantum_2014,cramer_repeated_2016}. Pioneering experiments have demonstrated codes that can detect but not correct errors \cite{Takita2017,linke2017fault,andersen2020repeated,marques2021logical}, as well as non-fault-tolerant quantum error correction protocols that could correct some, but not all, types of errors \cite{knill_benchmarking_2001,waldherr_quantum_2014,nigg_quantum_2014,cramer_repeated_2016, chen2021exponential,gong2021experimental,campagne2020quantum}. A recent experiment with trapped-ion qubits has demonstrated the fault-tolerant operation of an error correction code, albeit through destructive stabilizer measurements and post processing \cite{egan2020fault}.

In this work, we realize fault-tolerant encoding, gate operations and non-destructive stabilizer measurements for a logical qubit of a quantum error correction code. Our logical qubit is based on the 5-qubit code, and we use a total of seven spin qubits in a diamond quantum processor (Fig. \ref{fig1}). Fault-tolerance is made possible through the recently discovered paradigm of flag qubits \cite{chao2018,chamberland2018flag,chao2020PRXQ}. First, we demonstrate a novel fault-tolerant encoding protocol based on repeated multi-qubit measurements which herald the successful preparation of the logical state. Then, we realize a complete set of transversal single-qubit Clifford gates. Finally, we demonstrate stabilizer measurements on the logical qubit and include a flag qubit to ensure compatibility with fault-tolerance. Our stabilizer measurements are non-destructive, the post measurement state is available in real time, and we use feedforward based on the measurement outcomes. These results demonstrate the key principles of fault-tolerant quantum error correction in a solid-state spin-qubit processor.


\noindent\textbf{The logical qubit}\\
\noindent Stabilizer error correction codes use ancilla qubits to perform repeated stabilizer measurements that identify errors. A key requirement for fault-tolerance is to prevent errors on the ancilla qubits from spreading to the data qubits and causing logical errors (Fig. \ref{fig1}b) \cite{terhal_quantum_2015,chao2018}. The paradigm of flag fault-tolerance provides a solution with minimal qubit overhead \cite{chao2018, chamberland2018flag,chao2020PRXQ}. Ancilla errors that would cause logical errors are detected using additional flag qubits, so that they can be subsequently corrected (Fig. \ref{fig1}b). 

Our logical qubit is based on the 5-qubit code, the smallest distance-3 code which can correct any single-qubit error \cite{laflamme_1996,knill_benchmarking_2001}. Any logical state is a simultaneous $+1$ eigenstate of the four stabilizers $s_1 = XXYIY$, $s_2 = YXXYI$, $s_3 = IYXXY$, and $s_4 = YIYXX$, and the logical operators are $X_L = XXXXX$ and $Z_L = ZZZZZ$. Because any error on a single data qubit corresponds to a unique 4-bit syndrome, given as the eigenvalues of the stabilizers, arbitrary single-qubit errors can be identified and corrected. Combined with an ancilla qubit for stabilizer measurements and a flag qubit to capture harmful ancilla errors, this makes fault-tolerant error correction possible using 7 qubits in total \cite{chao2018}.

\noindent\textbf{System: spin qubits in diamond}\\
Our processor consists of a single nitrogen-vacancy (NV) centre and its surrounding nuclear spin environment at 4 Kelvin (Fig. 1a). These spins are high-quality qubits with coherence times up to seconds for the NV electron spin \cite{abobeih2018one} and minutes for the nuclear spins \cite{bradley2019ten}. The NV electron spin can be read out optically, couples strongly to all other spins, and is used as an ancilla qubit for stabilizer measurements (Methods) \cite{cramer_repeated_2016, bradley2019ten}. We use the intrinsic $^{14}$N nuclear spin as the flag qubit. In this device, 27 $^{13}$C nuclear-spin qubits and their lattice positions have been characterized, so that the 406 qubit-qubit interactions are known \cite{abobeih2019atomic}. Each $^{13}$C qubit can be controlled individually due to their distinct couplings to the NV electron spin (Methods). Here, we use five of the $^{13}$C spin qubits as the data qubits to encode the logical qubit. 

A challenge for controlling such a quantum processor is that the spins continuously couple to each other. We realize selective control gates through various echo sequences that isolate interactions between the targeted spins, while also protecting them from environmental decoherence. For all two-qubit gates, we use previously developed electron-nuclear gates, which are based on decoupling sequences on the electron spin (Methods) \cite{bradley2019ten}. Furthermore, we introduce interleaved and asynchronous echo stages that cancel unwanted couplings between the data qubits (Methods). These additional echo stages are essential for the relatively long gate sequences realized here.

\begin{figure}
\centering
\includegraphics[width=0.45\textwidth]{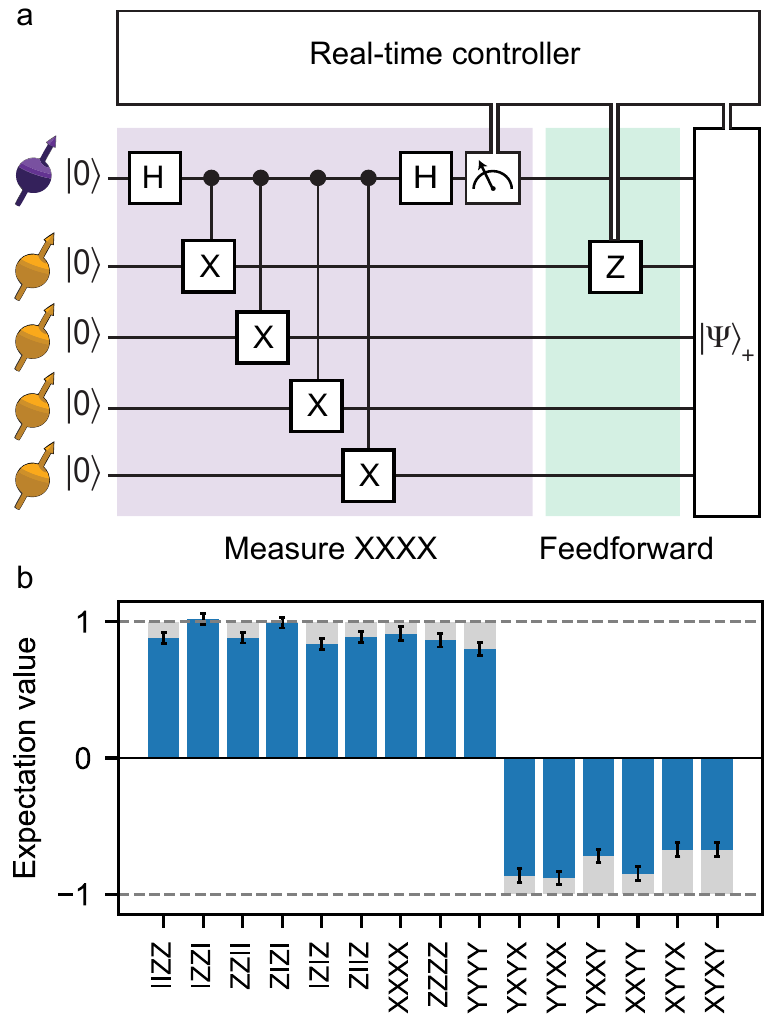}
\caption{Non-destructive stabilizer measurements with real-time feedforward. a) Circuit diagram for the deterministic preparation of a 4-qubit GHZ entangled state ($\ket{\Psi}_{+} =\ket{0000}+\ket{1111}/{\sqrt{2}}$) using a measurement of the stabilizer $XXXX$. b) Measured expectation values of the 15 operators that define the ideal state. The obtained fidelity with the target state is 0.86(1), confirming multi-partite entanglement. Grey bars show the ideal expectation values. Error bars are one standard deviation in all figures.}
\label{fig2}
\end{figure}

\noindent\textbf{Non-destructive stabilizer measurements}\\
We start by demonstrating non-destructive 4-qubit stabilizer measurements with real-time feedforward operations based on the measurement outcomes (Fig. 2). Despite the central role of such measurements in many error-correction codes \cite{gottesman_stabilizer_1997,preskill_reliable_1998,aliferis2005quantum,terhal_quantum_2015}, including the 5-qubit code, the Steane code and the surface code \cite{terhal_quantum_2015}, experimental implementations with feedforward have remained an outstanding challenge.

We benchmark the measurement by using it to deterministically create a 4-qubit entangled state. We prepare the state $\ket{0000}$ and measure the operator $XXXX$. This projects the qubits into the GHZ state $\ket{\Psi}_{\pm} = (\ket{0000}\pm\ket{1111})/{\sqrt{2}}$, with the sign determined by the measurement outcome. We process the measurement outcomes in real time using a microprocessor and apply the required correction to deterministically output the state $\ket{\Psi}_{+}$ with a fidelity of $0.86(1)$. Because this result is obtained without any post-selection, it highlights that the post-measurement state is available for all measurement outcomes, satisfying one of the key requirements for error correction. 

\begin{figure*}
\centering
\includegraphics[width= \textwidth]{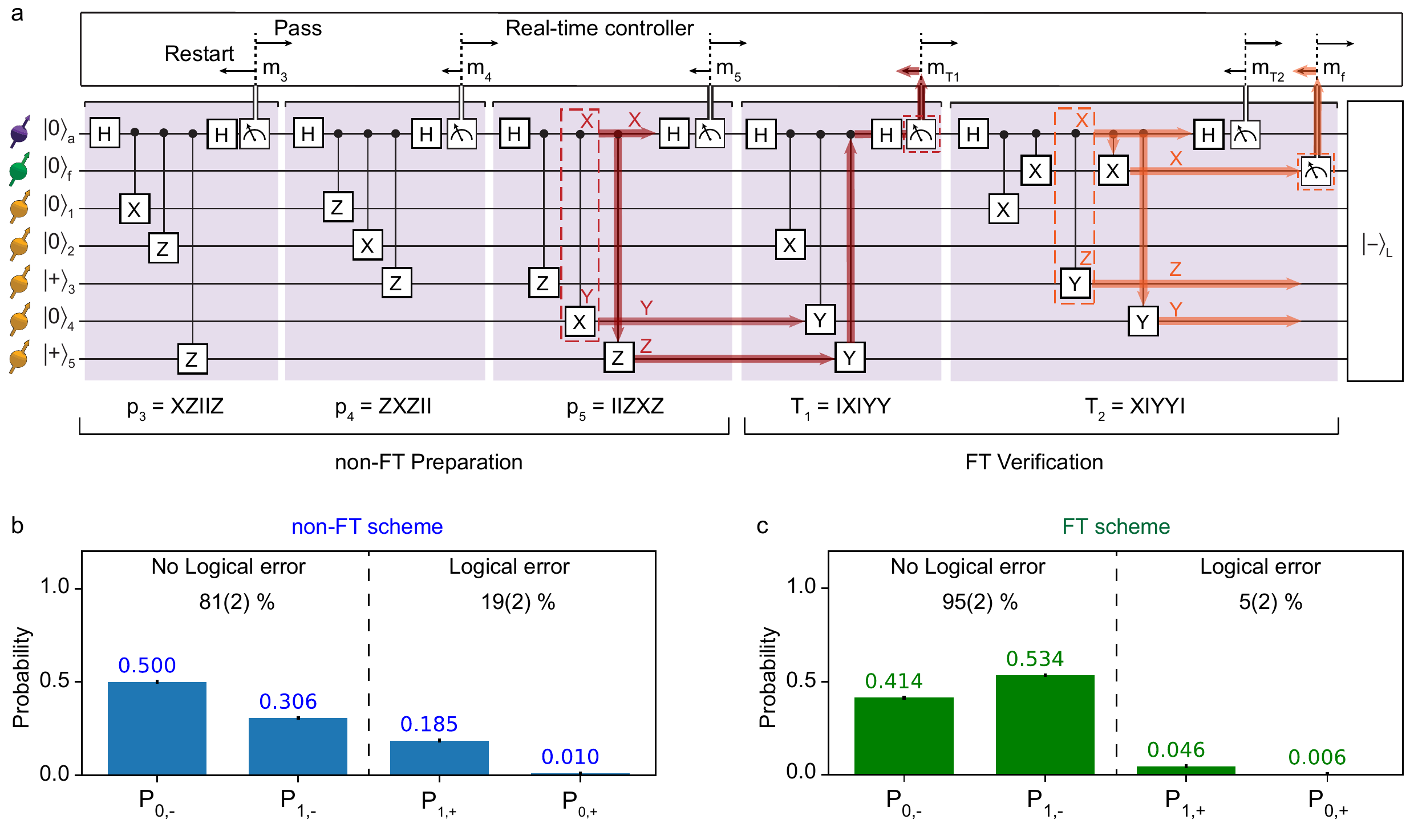}
\caption{Fault-tolerant encoding of the logical qubit. a) Encoding circuit. The first stage prepares $\ket{-}_L$ non-fault-tolerantly (`non-FT preparation') by starting with $\ket{00+0+}$ (an eigenstate of $p_1$, $p_2$) and measuring the logical operators $p_3,p_4,p_5$. The second `FT verification' stage consists of two stabilizer measurements $T_1 = p_2 p_4 p_5$ , $T_2 = p_1 p_3 p_5$ and a flag qubit measurement. Echo sequences are inserted between the measurements to decouple the qubits (not shown, see Supplementary Fig. 5). Successful preparation is heralded by satisfying a set of conditions for the measurement outcomes (see main text). Red: an example of an ancilla fault (an XY error in a two-qubit gate) that would propagate to a logical error but is detected by the $T_1$ verification step. Orange: an example of a single fault in the verification stage that would propagate into a logical error, but is detected by the flag qubit. b,c) Probabilities to obtain the desired logical state $\ket{-}_L$ without error ($P_{0,-}$) or with a single-qubit Pauli error ($P_{1,-}$), and the probabilities to obtain the opposite logical state $\ket{+}_L$ with zero ($P_{0,+}$) or with a single-qubit Pauli error ($P_{1,+}$). Note that $P_{1,\pm}$ are summed over all 15 possible errors. These $32$ states are orthogonal and span the full 5-qubit Hilbert space.} 
\label{fig3}
\end{figure*}
 

\noindent\textbf{Fault-tolerant encoding}\\
To prepare the logical qubit, we introduce a novel scheme that uses repeated stabilizer measurements and a flag qubit to herald successful preparation (Fig. 3a). In particular, we demonstrate the preparation of the logical state $\ket{-}_L = \frac{1}{\sqrt{2}}(\ket{0}_L-\ket{1}_L)$. This state is the unique $+1$ eigenstate of 5 independent weight-3 logical $-X$ operators namely $p_1 = IZXZI$, $p_2 = ZIIZX$, $p_3= XZIIZ$, $p_4= ZXZII$, $p_5= IIZXZ$. Therefore, one can prepare $\ket{-}_L$ by initializing the data qubits into the product state $\ket{00+0+}$, which is an eigenstate of $p_1$ and $p_2$, and subsequently measuring $p_3$ to $p_5$ (Fig. 3a). This preparation scheme is not fault-tolerant because faults involving the ancilla qubit can cause weight-2 errors, which can result in logical errors (Fig. 3a). We refer to these steps as the non-FT encoding scheme.

We make the preparation circuit fault-tolerant by adding two stabilizer measurements, $T_1 = p_2\cdot p_4\cdot p_5 = IXIYY$  and  $T_2 = p_1\cdot p_3\cdot p_5 = XIYYI $ with a flag qubit (Fig. 3a). Successful preparation is heralded by the following conditions: (1) the measurement outcomes of $T_1$ and $T_2$ are compatible with the measurement outcomes $m_i$ of the logical operators $p_i$, i.e.  $m_{T_1} = m_2\times m_4\times m_5$ and $m_{T_2} = m_1\times m_3\times m_5$; (2) the flag is not raised
(i.e., the flag qubit is measured to be in $\ket{0}$). 

Otherwise the state is rejected. Further details and a proof of the fault-tolerance of this scheme are given in the Supplementary Information. We refer to this preparation as the FT encoding scheme. 

To reduce the impact of ancilla measurement errors \cite{cramer_repeated_2016, van2019multipartite}, we additionally require all stabilizer measurement outcomes to be $+1$ (i.e., the NV electron spin is measured to be in $\ket{0}$). These outcomes are more reliable (see Methods) \cite{bradley2019ten}, increasing the fidelity of the state preparation, at the cost of a lower success probability (Supplementary Table 1). 

We compare the non-FT and FT encoding schemes. We define the logical state fidelity $F_{L}$ as (see Methods)
\begin{equation}\label{eq:logical_fidelity}
    F_{\rm L} = \sum_{E\in \mathcal{E}} \Tr( E\ket{-}_L\bra{-}_LE\cdot \rho),
\end{equation}
where $\rho$ is the prepared state, and $\mathcal{E} = \{I, X_i, Y_i, Z_i,\, i = 1, 2, \cdots,  5\}$ is the set of all single-qubit Pauli errors. The fidelity $F_{L}$ gives the probability that there is at most a single-qubit error in the prepared state, i.e. there is no logical error. We characterize the prepared state by tomographically measuring the 31 operators that define the target state (Supplementary Fig. 7 and Methods). We find that the FT encoding scheme ($F_{L} = 95(2) \%$) outperforms the non-FT scheme ($F_{L} = 81 (2) \%$).

To understand this improvement, we analyze the underlying error probability distributions (Figs. 3b,c). For the 5-qubit code, the $\ket{-}_L$ state plus any number of Pauli errors is equivalent to either $\ket{-}_L$ with at most one Pauli error (no logical error), or to $\ket{+}_L$ with at most one Pauli error (a logical error). We calculate the overlaps between the prepared state and those states. The results show that the FT scheme suppresses logical errors, consistent with fault-tolerance preventing single faults propagating to multi-qubit errors. The overall logical state fidelity $F_{L}$ is improved, despite the higher probability of single-qubit errors due to the increased complexity of the sequence.
 
 \begin{figure}
\centering
\includegraphics[scale=1.0]{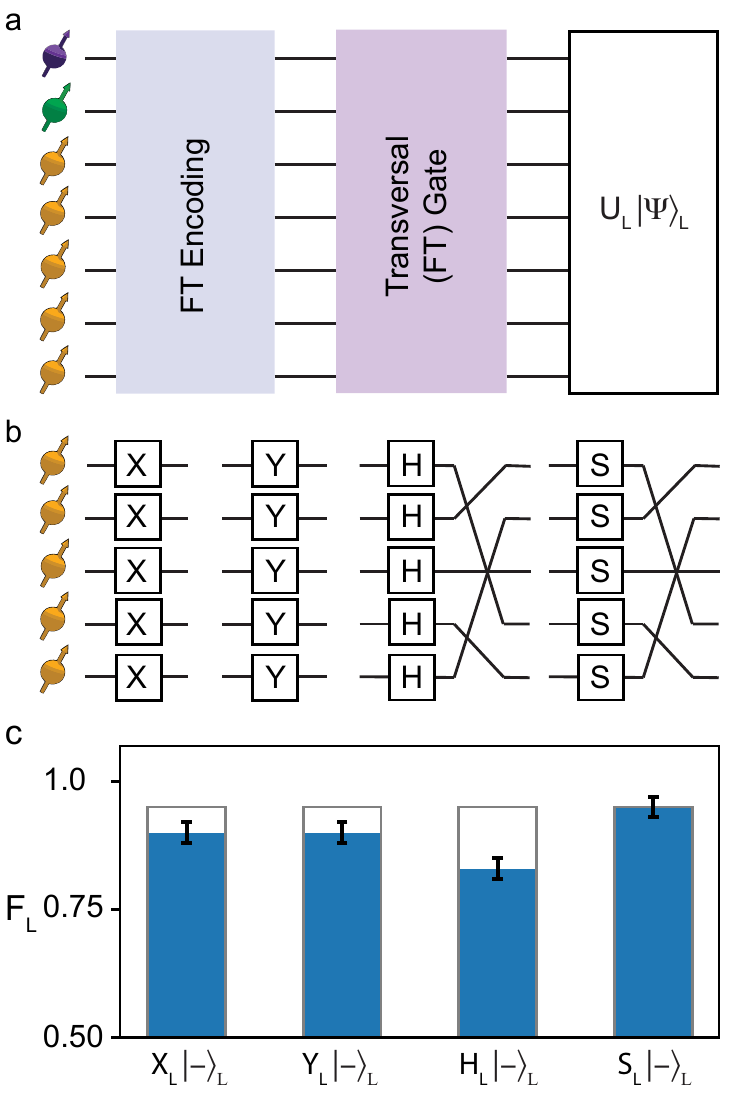}
\caption{Fault-tolerant gates on the logically-encoded qubit. a) We apply transversal logical gates on the encoded state $\ket{-}_L$ and measure the resulting logical state fidelity $F_L$ (Eq.~\eqref{eq:logical_fidelity}) with respect to the targeted state. b) Logical $X_L$,  $Y_L$, $H_L$ (Hadamard) and $S_L$ ($\pi/2$ rotation around the z-axis) are realized by five single-qubit gates. For $H_L$ and $S_L$ this is followed by a permutation of the qubits by relabelling them. c) Grey bars: logical state fidelity when compiling the logical gates with subsequent operations ($0.95(2)$ for all gates). Blue bars: logical state fidelities after physically applying the transversal logical gates (0.90(2),0.90(2),0.83(2),0.95(2) for $X_L, Y_L, H_L$, and $S_L$ respectively).} 
\label{fig4}
\end{figure}


\noindent\textbf{Fault-tolerant logical gates}\\
The 5-qubit code supports a complete set of transversal single-qubit Clifford gates, which are naturally fault-tolerant \cite{gottesman_stabilizer_1997,Yoder_2016}. We apply four transversal logical gates to $\ket{-}_L$ (Fig. 4): $X_L = X_1 X_2 X_3 X_4 X_5$,  $Y_L = Y_1 Y_2 Y_3 Y_4 Y_5$, the Hadamard gate $H_L = P_{\pi} H_1 H_2 H_3 H_4 H_5$, and the phase gate $S_L = P_{\pi} S_1 S_2 S_3 S_4 S_5$, where $P_{\pi}$ is a permutation of the data qubits (see Fig. 4b) \cite{gottesman_stabilizer_1997,Yoder_2016}. These permutations are fault-tolerant because we realize them by relabelling the qubits rather than by using SWAP gates \cite{Yoder_2016}. 

Our control system performs the underlying single-qubit gates by tracking basis rotations and compiling them with subsequent gates or measurements (Methods). In the sequence considered here (Fig. 4a), such compilation does not increase the physical operation count, and there is no reduction of fidelity (Fig. 4c). For comparison, we also implement the `worst-case' scenario where the logical gates are applied physically (Fig. 4c). This includes 5 single-qubit gates and the corresponding additional echo sequences between the state preparation and the measurement stage. Together, the demonstrated transversal logical gates enable the fault-tolerant preparation of all six eigenstates of the logical Pauli operators.  

\begin{figure}
\centering
\includegraphics[scale=0.7]{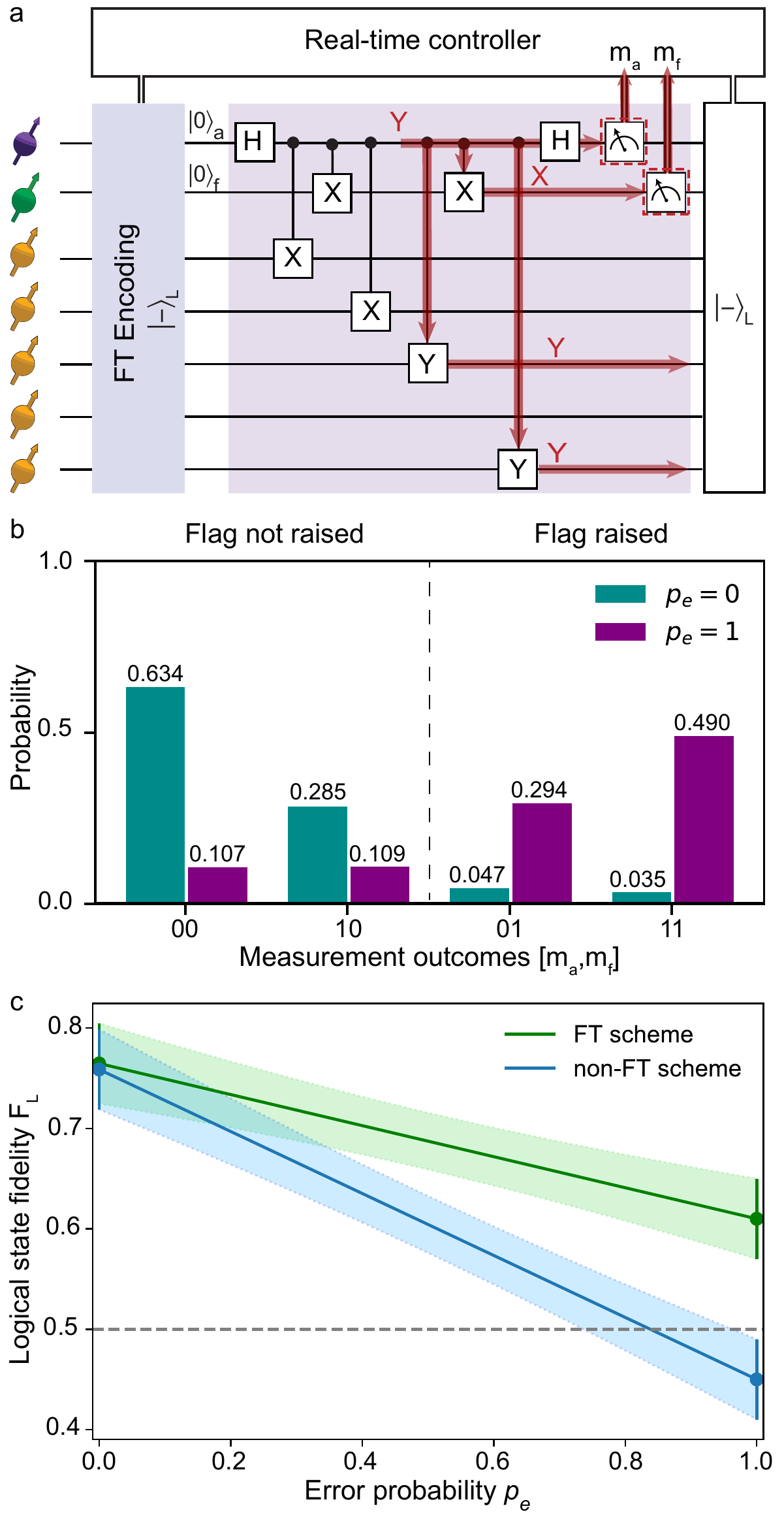}
\caption{Fault-tolerant stabilizer measurement. a) Circuit diagram to measure the stabilizer $XXYIY$ on the encoded state. As an example to illustrate the compatibility with fault-tolerance we insert a $Y$ error on the ancilla qubit. This error will propagate to the 2-qubit error $Y_3Y_5$ on the data qubits, which leads to a logical $Z$ error. However, because the error also triggers the flag qubit it can be accounted for (Methods). b) Probability of the measurement outcomes of the ancilla ($m_a$) and flag ($m_f$) qubits when inserting ($p_e = 1$) or not inserting ($p_e = 0$) the $Y$ error on the ancilla qubit. The results show that the flag qubit successfully detects this error. c) Logical state fidelity $F_L$ after the stabilizer measurement as function of the error probability $p_e$. The non-FT case does not take the flag outcome into account. Values between $p_e = 0$ and $p_e = 1$ are calculated as weighted sums (Methods).}
\label{fig5}
\end{figure}

\noindent\textbf{Fault-tolerant stabilizer measurements}\\
Finally, we demonstrate and characterize a flagged stabilizer measurement on the encoded state (Fig. 5a). Such measurements are a primitive for fault-tolerant quantum error correction protocols \cite{chao2018}. To ensure that the measurement is compatible with fault-tolerance, the two-qubit gates are carefully ordered and a flag qubit is added to capture ancilla errors that can propagate to logical errors \cite{chao2018}. 

We prepare the logical state $\ket{-}_L$ and measure the stabilizer $s_1 = XXYIY$ (Fig. 5a). The resulting output consists of the post-measurement state and two classical bits of information from the ancilla and flag measurements (Fig. 5b). The logical state fidelity $F_L$ is given by the probability that the logical information can be correctly extracted (no logical error), when taking into account the flag measurement outcome. The interpretation of the error syndrome changes if the flag is raised (Methods). We find $F_L=0.77(4)$ for the post measurement state without any post-selection. Higher logical state fidelities can be obtained by post-selecting on favourable outcomes, but this is incompatible with error correction.      

To illustrate the benefit of the flag qubit we compare the logical state fidelities with and without taking the flag measurement outcome into account. Because ancilla errors that propagate to logical errors are naturally rare, no significant difference is observed (Fig. 5c). Therefore, we introduce a Pauli $Y$ error on the ancilla qubit (Fig. 5a). This error propagates to the 2-qubit error $Y_3Y_5$. For the case without flag information, this error causes a logical flip $Z_L$ (Methods), and the logical state fidelity drops below 0.5. In contrast, with the flag qubit, this non-trivial error is detected (Fig. 5b) and remains correctable, so that the logical state fidelity is partly recovered (Fig. 5c).


\noindent\textbf{Conclusion}\\
In conclusion, we have demonstrated encoding, gates and non-destructive stabilizer measurements for a logical qubit of an error correction code in a fault-tolerant way. Such fault-tolerant operations are a necessity for large-scale quantum computation in which error rates ultimately must be suppressed to extremely low levels. 

Future challenges are to perform complete quantum error correction cycles, encode multiple logical qubits, realize universal fault-tolerant gates, and ultimately suppress logical error rates exponentially below physical error rates. While the demonstrated operations are of high fidelity\textemdash the experiments consist of up to 40 two-qubit gates and 8 mid-circuit ancilla readouts (Fig.~5a)\textemdash improvements in both the fidelities and the number of qubits will be required. 

Improved gates might be realized through tailored optimal control schemes that leverage the precise knowledge of the system and its environment (Fig. 1a) \cite{dolde_high-fidelity_2014}. Coupling to optical cavities can further improve readout fidelities \cite{bhaskar2020experimental,awschalom2018quantum}. Scaling to large code distances and multiple logical qubits can be realized through already-demonstrated magnetic \cite{dolde_high-fidelity_2014} and optical \cite{pompili2021realization} NV-NV connections that enable modular, distributed, quantum computation based on the surface code and other error correction codes \cite{nickerson_topological_2013}. Therefore, our demonstration of the building blocks of fault-tolerant quantum error correction is a key step towards quantum information processing based on solid-state spin qubits.

\noindent\textbf{Methods}\\
\textbf{Sample.}
We use a naturally occurring NV centre in a homo-epitaxially chemical-vapor-deposition (CVD) grown diamond with a $1.1\%$ natural abundance of $^{13}$C and a $\langle 111 \rangle$ crystal orientation (grown by Element Six). A solid-immersion lens is used to enhance the photon-collection efficiency \cite{Robledo_Nature2011}. The NV center has been selected for the absence of $^{13}$C spins with hyperfine couplings $>$ 500 kHz. These experiments are performed at a temperature of 4 K where the electron-spin relaxation is negligible ($T_1=3.6(3) \times 10^3\,$s \cite{abobeih2018one}). 

\noindent\textbf{Qubits and coherence times.}
The NV electron-spin ancilla qubit is defined between the states $m_s=0$ ($\ket{0}$) and $m_s=-1$ ($\ket{1}$). The NV electron-spin coherence times are $T_2^* = 4.9(2)\,\mu$s, $T_2 = 1.182(5)\,$ms, and up to seconds under dynamical decoupling \cite{abobeih2018one}. The $^{14}$N nuclear-spin flag qubit is defined between the states $m_I=0$ ($\ket{0}$) and $m_I=-1$ ($\ket{1}$). The $^{13}$C nuclear-spin data qubits in this device have been characterized in detail in previous work (Fig. 1a) \cite{abobeih2019atomic,bradley2019ten,jung2020deep}. See Supplementary Tables 2-5 for the hyperfine parameters, coherence times, and qubit-qubit interactions for the qubits used here. 

\noindent\textbf{Magnetic field.}
A magnetic field of $\sim403\,$G is applied using a room-temperature permanent magnet on a XYZ translation stage. This applied field lifts the degeneracy of the $m_s = \pm 1$ states due to the Zeeman term (Supplementary Information). We stabilise the magnetic field to $<$ 3 mG using temperature stabilisation and an automatic re-calibration procedure (every few hours). We align the magnetic field along the NV axis using thermal echo sequences with an uncertainty of $0.07$ degrees in the alignment \cite{abobeih2019atomic}. 

\noindent \textbf{Single- and two-qubit gates.}
Single-qubit gates and echo pulses are applied using microwave pulses for the NV electron spin ($m_s = 0 \leftrightarrow m_s = -1$ transition, Hermite pulse shapes \cite{abobeih2018one,warren1984effects}, Rabi frequency of $\sim 15\,$MHz) and using radio-frequency pulses for the $^{13}$C spin qubits (error function pulse shapes \cite{bradley2019ten}, typical Rabi frequency of $\sim 500\,$Hz) and the $^{14}$N spin qubit (error function pulse shapes, Rabi frequency $\sim 2\,$kHz). The $^{13}$C spin qubits (data qubits) are distinguishable in frequency due to their hyperfine coupling to the NV electron spin (Supplementary Table 2).

Electron-nuclear two-qubit gates are realised using two different gate designs, depending on the properties of the targeted nuclear spin. For data qubits 1,2,4,5, two-qubit gates are realized through dynamical decoupling sequences of $N$ equally spaced $\pi$-pulses on the electron spin of the form ($\tau_r -\pi - \tau_r)^{N}$ \cite{taminiau_universal_2014,bradley2019ten}. This design requires a significant hyperfine component perpendicular to the applied magnetic field \cite{taminiau_universal_2014}. For data qubit 3 and the flag qubit (the $^{14}$N spin) the perpendicular hyperfine coupling is small and we perform two-qubit gates by interleaving the dynamical decoupling sequence with RF pulses \cite{bradley2019ten}. Both gate designs simultaneously decouple the NV electron spin from the other qubits and the environment \cite{bradley2019ten}. The parameters and fidelities for the two-qubit gates are given in Supplementary Table 4

\noindent\textbf{Compilation of gate sequences.} Our native two-qubit gates are electron-controlled nuclear-spin rotations and are equivalent to the CNOT gate up to single-qubit rotations (Supplementary Fig.~3). To implement the sequences shown in the figures, we first translate all gates into these native gates and compile the resulting sequence. 

\noindent\textbf{Echo sequences for the data qubits.} To mitigate decoherence of the data qubits due to their spin environment, we use echo sequences that are interleaved throughout the experiments. These echo sequences ensure that the data qubits rephase each time they are operated on. Additionally, the sequence design minimizes the time that the ancilla electron spin is idling in superposition states, which are prone to dephasing. We use two echo stages between stabilizer measurements, as well as before and after the logical gates of Fig. 4, which provides a general and scalable solution for the timing of all gates and echoes (Supplementary Information). 

An additional challenge is that, due to the length of the sequences (up to 100 ms), we need to account for the small unwanted interactions between the nuclear-spin data qubits. The measured coupling strengths show that the strongest couplings are between qubits 3 and 2 (16.90(4) Hz, and between qubits 3 and 5 (12.96(4) Hz) (Supplementary Table 5) \cite{abobeih2019atomic}). Such interactions can introduce correlated two-qubit errors that are not correctable in the distance-3 code considered here, which can only handle single-qubit errors in the code block. 

To mitigate these qubit-qubit couplings, we decouple qubit 3 asynchronously from the other qubits (Supplementary Fig.~5). Ultimately, such local correlated errors can be suppressed entirely by larger distance codes.

\noindent\textbf{Real-time control and feedforward operations.} Real-time control and feedforward operations are implemented through a programmable microprocessor (Jaeger ADwin Pro II) operating on microsecond timescales. The microprocessor detects photon events coming from the detectors, infers the measurement outcomes, and controls both the subsequent sequences in the arbitrary waveform generator (Tektronix AWG 5014c) and the lasers for the ancilla qubit readout. The precise timing for quantum gates ($1$ nanosecond precision) is based on the clock of the arbitrary waveform generator. Additionally, the microprocessor operates various control loops that prepare the NV center in the negative charge state, on resonance with the lasers, and in the focus of the laser beam (see Supplementary Information).    

\noindent\textbf{Readout of the ancilla qubit.}
The electron spin (ancilla qubit) is read out by resonantly exciting the $m_s=0$ to $E_x$ optical transition \cite{Robledo_Nature2011}. For one or more photons detected we assign the $m_s=0$ outcome, for zero photons we assign $m_s = \pm1$. The single-shot readout fidelities are $F_0 = 90.5(2) \%$ and $F_1 = 98.6(2) \%$ for $m_s=0$ and $m_s=-1$, respectively (average fidelity: $94.6(1)\%$). 

Uncontrolled electron-spin flips in the excited state cause dephasing of the nuclear spins through the hyperfine interaction. To minimize such spin flips we avoid unnecessary excitations by using weak laser pulses, so that a feedback signal can be used to rapidly turn off the laser upon detection of a photon (within 2 $\mu$s). The resulting probability that the electron spin is in state $m_s=0$ after correctly assigning $m_s=0$ in the measurement is 0.992 \cite{cramer_repeated_2016}.

For measurements that are used for heralded state preparation, i.e. where we only continue upon a $m_s=0$ outcome (see e.g. Fig. 3), we use shorter readout pulses. This improves the probability that a $m_s=0$ outcome correctly heralds the $m_s=0$ state, at the cost of reduced success probability (see Supplementary Table 1).

\noindent\textbf{System preparation and qubit initialization.} At the start of the experiments we first prepare the NV center in its negative charge state and on resonance with the lasers. We then initialize the NV electron spin in the $m_s = 0$ state through a spin pumping process (Fidelity $> 99.7 \%$) \cite{Robledo_Nature2011}. We define the electron-spin qubit between the states $m_s=0 (\ket{0})$ and $m_s = -1 (\ket{1})$. We initialize the data qubits through SWAP sequences (Supplementary Fig.~4) into $\ket{0}$, and subsequent optical reset of the ancilla qubit (initialization fidelities $96.5 \% - 98.5 \%$, see Supplementary Table 4). The flag qubit is initialized through a projective measurement that heralds preparation in $\ket{0}$ (initialization fidelity $99.7 \%$). Other product states are prepared by subsequent single qubit gates.

\noindent\textbf{Final readout of the data qubits.}
Tomographically measuring single- and multi-qubit operators of the data qubits is performed by mapping the required correlation to the ancilla qubit (through controlled rotations) and then reading out the ancilla qubit \cite{cramer_repeated_2016}. In order to provide best estimates for the measurements, we correct the measured expectation values (Fig.~2 and Supplementary Figs.~6,7) for infidelities in the readout sequence, see Bradley et al. \cite{bradley2019ten} for the correction procedure.

\noindent \textbf{Assessing the logical state fidelity.} The logical state fidelity $F_L$ is defined in Eq.~(1) and gives the probability that the state is free of logical errors.  Said differently, $F_L$ is the fidelity with respect to the ideal 5-qubit state after a round of perfect error correction, or the probability to obtain the correct outcome in a perfect fault-tolerant logical measurement. While fault-tolerant circuits for logical measurement exist \cite{chao2018}, we do not experimentally implement these here. Instead, we extract $F_L$ from a set of tomographic measurements, as described in the following using $\ket{-}_L$ as an example.

The logical state $\ket{-}_L$ is the unique simultaneous eigenstate of the 5 weight-3 operators $p_i$ with eigenvalue $+1$. We can thus describe the state $E\ket{-}_{L}$ (with $E$ a Pauli error) as the projector
\begin{align*}
E\ket{-}_{L}\bra{-}_{L}E = \prod_{i=1}^5 \left( \frac{1+m_i p_i}{2} \right),
\end{align*}
where $m_i =\pm 1$ is the measurement outcome of $p_i$, and $m_i = -1$ when $E$ anti-commutes with $p_i$. This projector can be expanded as a summation of 31 multi-qubit Pauli operators (including a constant), which are listed in Supplementary Fig.~7.
The logical state fidelity $F_L$ in Eq.~(\ref{eq:logical_fidelity}) can then be written as
\begin{equation}\label{eq:FL_singleQubitCorrection}
\begin{split}
  F_L =& \sum_{E\in \mathcal{E}} \Tr( E\ket{-}_L\bra{-}_L E\, \rho)\\
	=&\, \frac{1}{2} + \frac{1}{8} ( \langle IZXZI\rangle  + \langle ZIIZX\rangle  + \langle XZIIZ\rangle\\
	&\quad  + \langle ZXZII\rangle + \langle IIZXZ\rangle  + \langle YIXIY\rangle\\
	&\quad   + \langle IYYIX\rangle  + \langle XIYYI\rangle 
	+ \langle IXIYY\rangle\\
	&\quad + \langle YYIXI\rangle  + \langle ZZYXY\rangle  + \langle YXYZZ\rangle  \\
	&\quad + \langle ZYXYZ\rangle  + \langle XYZZY\rangle  + \langle YZZYX\rangle\\
	&\quad  + \langle XXXXX\rangle ).
\end{split}
\end{equation}  
Here $\mathcal{E} = \{I, X_i, Y_i, Z_i,\, i = 1, 2, \cdots 5\}$ is the set of correctable errors for the 5-qubit code. To obtain $F_L$ experimentally, we measure this set of expectation values.

\noindent\textbf{Logical state fidelity with flag.} If the flag in the circuit in Fig.~\ref{fig5}a is not raised, then a cycle of error correction would correct any single-qubit error on a logical state. The logical state fidelity is then given by Eq.~(\ref{eq:logical_fidelity}), which we now refer to as $F_L^{\rm not \ raised}$. A raised flag leads to a different interpretation of the error syndrome (Supplementary Table 7) \cite{chao2018}.

For example, the $Y$ error on ancilla in Fig.~\ref{fig5}a leads to the output state $Y_3Y_5\ket{-}_L$, for which the eigenvalues of $s_1 = XXYIY, s_2 = YXXYI, s_3 = IYXXY, s_4 = YIYXX$ give the syndrome $[+1, -1, -1, -1]$. Without flag, the corresponding single-qubit recovery is $Z_4$, which changes the syndrome back to all $+1$ (Supplementary Table 7). This recovery leads to the remaining error $Y_3Z_4Y_5$, which is a logical $Z$ error. However, taking the flag measurement outcome into account, the syndrome is interpreted differently and the recovery is $Y_3Y_5$ so that no error is left (Supplementary Table 7). 

For the cases where the flag is raised, the logical state fidelity with respect to $\ket{-}_L$ is now given by:
\begin{equation}\label{eq:p_good}
\begin{split}
	F_{\rm L}^{\rm raised} =& \sum_{E\in \mathcal{E}'} \Tr( E\ket{-}_L\bra{-}_L E\cdot \rho)\\
	=&\, \frac{1}{2} + \frac{1}{32} \big( 6\langle IIZXZ\rangle + 6\langle ZXZII\rangle + 6\langle YYIXI\rangle \\&\quad  - 2 \langle ZIIZX\rangle + 6\langle IXIYY \rangle 
 	+ 2 \langle YZZYX \rangle\\&\quad + 2 \langle XYZZY\rangle - 2\langle IZXZI \rangle + 2\langle ZYXYZ\rangle\\&\quad + 6\langle XIYYI\rangle +2\langle YXYZZ\rangle +2\langle ZZYXY\rangle\\&\quad + 6\langle IYYIX\rangle - 2\langle YIXIY\rangle + 2\langle XXXXX\rangle\\&\quad - 2\langle XZIIZ\rangle\big)
\end{split}
\end{equation} 
with $\mathcal{E}'$ another set of correctable errors 
\begin{equation}
\begin{split}
    \mathcal{E}' = \{I, X_1, X_3Y_5, Z_1,& X_2, Y_2, Z_3Y_5, X_1Y_2, Y_3, Z_3,\\& X_4, Y_4, Y_3Y_5, X_5, Y_5, X_1Z_2 \}.
\end{split}
\end{equation}
A detailed derivation for this set of errors and their corresponding syndromes are given in the Supplementary Information.

The logical state fidelity after the stabilizer measurement (Fig. 5) is calculated as the weighted sum of the fidelities conditioned on the two flag outcomes:
\begin{equation}
F_L = p_{\rm f}\, F_L^{\rm raised} + (1-p_{\rm f})\,F_L^{\rm not \ raised},
\end{equation}
with $p_{\rm f}$ the probability that the flag is raised and $F_L^{\rm raised}$ and $F_L^{\rm not\ raised}$ are defined above.

Finally, to construct the logical state fidelity as a function of $p_e$ (Fig. 5c), we measure $F_L$ with ($p_e = 0$) and without ($p_e = 1$) the ancilla error and calculate the outcomes for other error probabilities $p_e$ from their weighted sum:
\begin{equation}
F_L(p_e) = (1-p_e)F_{L}(p_e = 0) + p_eF_L(p_e = 1)
\end{equation}

\noindent\textbf{Error distribution in the prepared state.}
The overlaps between the prepared state $\rho$ and the state $E\ket{-}_L$ with $E$ identity or a single-qubit error are written as $P_{0,-}$ and $P_{1,-}$, respectively. These correspond to the cases that there is no logical error. The overlaps between the prepared state $\rho$ and the state $E\ket{+}_L$ with $E$ identity or a single-qubit error are written as $P_{0,+}$ and $P_{1,+}$, respectively. In these cases there is a logical error. These overlaps are shown in Fig.~\ref{fig3}b,c and calculated as ($\alpha=\pm$)

\begin{equation}
  P_{0,\alpha} = \Tr( \ket{\alpha}_L \bra{\alpha}_L \cdot \rho), 
\end{equation}

\begin{equation}
    P_{1,\alpha} = \sum_{E\in \mathcal{E}} \Tr( E\ket{\alpha}_L \bra{\alpha}_L E\cdot \rho
    ). 
\end{equation}

\noindent\textbf{Acknowledgements.} This work was supported by the Netherlands Organisation for Scientific Research
(NWO/OCW) through a Vidi grant and the Quantum Software Consortium programme (Project No. 024.003.037/3368). This project has received funding from the European Research Council (ERC) under the European Union’s Horizon 2020 research and innovation programme (grant agreement No. 852410). We gratefully acknowledge support from the joint research program “Modular quantum computers” by Fujitsu Limited and Delft University of Technology, co-funded by the Netherlands Enterprise Agency under project number PPS2007. This project (QIA) has received funding from the European Union’s Horizon 2020 research and innovation programme under grant agreement No 820445. This work was supported by a QuantERA grant for the QCDA consortium. \\

\noindent\textbf{Note added.} While finalizing this manuscript two related pre-prints appeared that demonstrate destructive stabilizer measurements with a flag qubit \cite{hilder2021fault}, and flag fault-tolerant quantum error correction \cite{ryan2021realization} with trapped-ion qubits.

\noindent\textbf{Data availability.} \\
The data that support the findings of this study are available from the corresponding author upon request.

\noindent\textbf{Author contributions.} \\
MHA, YW, and THT devised the experiments. MHA performed the experiments and collected the data. YW and BMT developed the fault-tolerant preparation scheme and its analysis. MHA, YW, JR, BMT, and THT analyzed the data. MHA, JR, SJHL and CEB prepared the experimental apparatus. MM and DJT grew the diamond sample. MHA, YW and THT wrote the manuscript with input from all authors. BMT and THT supervised the project.

\bibliographystyle{naturemag}
\bibliography{bibliography}

\end{document}



\begin{center}
\textbf{\Large Supplementary Information for ``Fault-tolerant operation of a logical qubit in a diamond quantum processor''}
\end{center}

\tableofcontents
\newpage

\section{System Hamiltonian}
The Hamiltonian describing the physical system of an NV centre and the surrounding nuclear-spin qubits can be approximated as \cite{abobeih2021thesis}:
\begin{equation}
\label{system_hamiltonian}
    H = H_e + H_N + H_C + H_{eN} + H_{eC} + H_{CC}, 
\end{equation}
where $H_{e}$ is the NV electron spin Hamiltonian, $H_N$ is the Hamiltonian of the $^{14}$N nuclear spin, $H_c$ is the Hamiltonian of the $^{13}$C spins, $H_{eN}$ is the hyperfine interaction between the electron spin and the host $^{14}$N nuclear spin, $H_{eC}$ is the hyperfine interaction between the electron spin and the $^{13}$C nuclear spins, and $H_{CC}$ is the coupling between the nuclear-spin qubits.\\

\noindent{\bf Electron spin.}  The NV electron spin ground state is a spin triplet ($S = 1$) and its Hamiltonian can be approximated as (neglecting second order spin-orbit coupling) \cite{doherty_theory_2012}:

\begin{equation}
    H_e =  \Delta_\text{ZFS} S_z^2 + \gamma_e \bf B \cdot \bf S, 
\end{equation}
where $\Delta_\text{ZFS} \approx 2.88$ GHz is the zero field splitting, $\gamma_e \approx 2.8$ MHz/G  is the electron gyromagnetic ratio, ${\bf B} = (B_x, B_y, B_z)$ is the magnetic field vector, and ${\bf S} = (S_x, S_y, S_z)$ are the electron spin-1 operators. The spin states $m_s=0$ and $m_s = \pm 1$ are split by $\Delta_\text{ZFS}$ at zero applied magnetic field. In this work, we apply an external magnetic field of  $\sim403\,$G which lifts the degeneracy of the $m_s = \pm 1$ states due to the Zeeman term. We define our electron-spin qubit between the states $m_s=0$ ($\ket{0}$) and $m_s=-1$ ($\ket{1}$).\\

\noindent{\bf Nitrogen spin.} The $^{14}$N nuclear-spin Hamiltonian $H_N$ and its interaction with the electron spin $H_{eN}$ can be written as:

\begin{equation}
    H_N +H_{eN} =  -Q_N I_{N,z}^2 + \gamma_N {\bf B} \cdot {\bf I}_N + {\bf S} \cdot {\bf A}_N \cdot {\bf I}_{N}, \\
\end{equation}
where $Q_N \approx 4.98$ MHz is the quadrupole splitting (splitting the energy of the nitrogen-spin states $m_I=0$ and $m_I =\pm 1$ at zero magnetic field) \cite{doherty_theory_2012}, $\gamma_N = 0.3077$ kHz/G  is the nitrogen-spin gyromagnetic ratio, ${\bf I}_{N} = (I_{N,x}, I_{N,y}, I_{N,z})$ are the spin-1 operators for the $^{14}$N nuclear spin, and ${\bf A}_N$ is the hyperfine tensor describing the electron-$^{14}$N interaction. Note that under the secular approximation, $H_{eN}$ can be approximated as $A_{\parallel} \hat{S}_{z}\hat{I}_{N,z}$, where $A_{\parallel} = 2.16$ MHz is the parallel component of the hyperfine interaction.\\

\noindent{\bf Carbon-13 spins.} 
The Hamiltonian for the $^{13}$C spins $H_c$ and their hyperfine interaction with the electron-spin $H_{ec}$ can be written as:

\begin{equation}
H_{C} + H_{eC} = \gamma_C \sum_{i} {\bf B}\cdot{\bf I}_{i} + \sum_{i} {\bf S} \cdot {\bf A}_{i} \cdot {\bf I}_{i},
\end{equation}
where  $\gamma_C = 1.0705 $ kHz/G  is the carbon-spin gyromagnetic ratio, ${\bf I}_{i} = (I_{i,x}, I_{i,y}, I_{i,z})$ are the spin-1/2 operators for the $^{13}$C nuclear spins, and ${\bf A}_{i}$ is the hyperfine tensor describing the electron-$^{13}$C interaction. Under the secular approximation, $H_{eC}$ can be approximated to $\sum_{i} A_{\parallel} \hat{S}_{z}\hat{I}_{i,z} + A_{\perp} \hat{S}_{z}\hat{I}_{i,x} $, where $A_{\parallel}$ and $A_{\perp}$ are the hyperfine interaction components parallel and perpendicular to the applied magnetic field.\\

\noindent{\bf Nuclear-nuclear interactions.}
In addition to their interactions with the NV electron spins, the nuclear-spin qubits interact with each other via a weak dipolar coupling. For the work presented here, $H_{CC}$ can be approximated as

\begin{equation}
H_{CC} = \sum_{i,j} C_{ij} \hat{I}_{i,z} \hat{I}_{j,z},
\end{equation}
where $C_{ij}$ are the dipolar spin-spin couplings, slightly modified due to the presence of the electron spin \cite{abobeih2019atomic}, and have been characterised for the system used here (Supplementary Table \ref{Tab:nuclear-nuclear couplings}).
\section{Experimental sequence}
\begin{figure}[!htb]
\centering
\includegraphics[width=\columnwidth,angle=0]{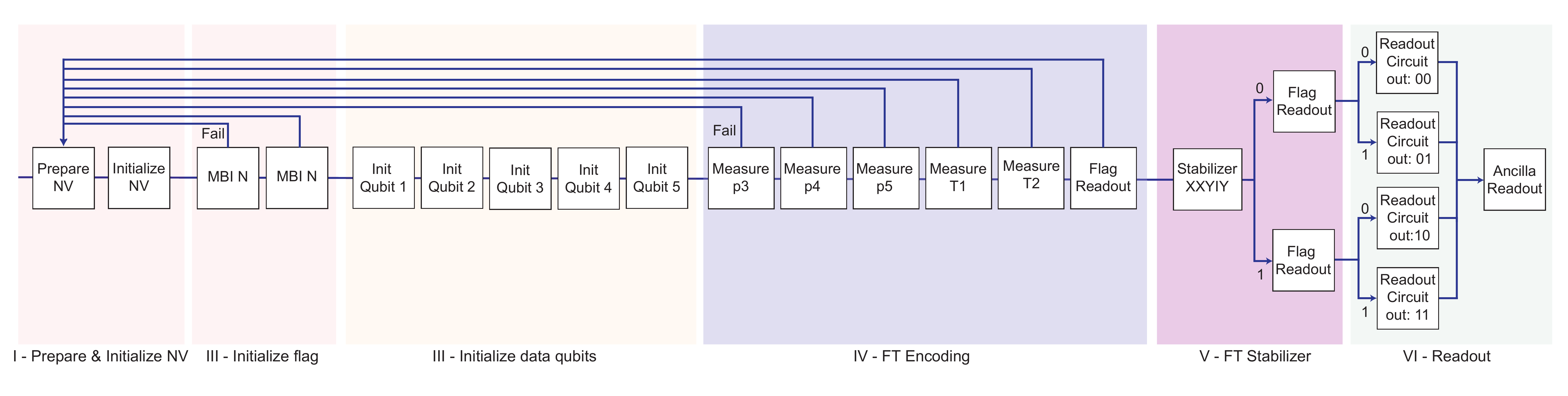}
\caption{Block diagram of the experimental sequence and logic. \textbf{I.} The NV center is prepared in the negative charge state and on resonance with the lasers, and then initialized into the $m_s = 0$ state through optical spin pumping \cite{Robledo_Nature2011}. See Supplementary Fig.~\ref{fig:NV_preparation} for details. \textbf{II.} The $^{14}$N nuclear-spin flag qubit is initialized using measurement-based initialization (MBI), which is performed twice to improve the initialization fidelity (Supplementary Fig.~\ref{fig:init_tomo}). \textbf{III.} The five data qubits are initialized sequentially by SWAP sequences that transfer the electron spin state ($\ket{0}$) onto the data qubits (Supplementary Fig.~\ref{fig:init_tomo}). \textbf{IV.} To fault-tolerantly encode the logical qubit, we measure the logical operators $p_3,p_4,p_5$, the verification operators $T_1 = p_2 p_4 p_5$, $T_2 = p_1 p_3 p_5$, and the flag qubit. \textbf{V.} After a successful logical state preparation, we measure the stabilizer $s_1 = XXYIY$ with a flag qubit measurement check. \textbf{VI.} Finally, to characterize the post measurement state and obtain the logical state fidelity, we perform multi-qubit readout of the data qubits by mapping the required correlation to the ancilla qubit and then reading out the ancilla qubit (Supplementary Fig.~\ref{fig:init_tomo}). The branches and return arrows in the diagram indicate real-time decisions and feedforward based on the measurement outcomes. Optional extra blocks for logical gates (Fig.~4 in the main text) are not shown.}
\label{fig:Block_diagram}
\end{figure}

In this section we describe the experimental sequence (Supplementary Fig.~\ref{fig:Block_diagram}) and the implementation of the presented quantum circuits using our native gates in more detail. The NV electronic spin is chosen as the ancilla qubit: first, it can be initialized and non-destructively read out by optical means; second, it can be used to directly implement two qubit gates with all of the $^{13}$C nuclear-spin qubits as well as the $^{14}$N nuclear-spin qubit. We use the $^{14}$N nuclear spin as the flag ancilla, and five $^{13}$C nuclear spins as the data qubits.

Supplementary Fig.~\ref{fig:Block_diagram} provides an overview of the typical blocks in the experimental sequences. The qubit initialization and procedure is described in more detail in Supplementary Fig.~\ref{fig:init_tomo}, the encoding of the logical qubit and the layout of the data qubit echo sequences is given in Supplementary Fig.~\ref{fig:Encoding_gates}, the data qubit readout is given in Supplementary Fig.~\ref{fig:init_tomo}. Below we describe the native two-qubit gates in the system, the translation and compilation of the main text circuits into the experimental gate sequences and the implementation of echo sequences to decouple the qubits from each other and from the environment.

\clearpage

\subsection{NV preparation and initialization}

\begin{figure}[!htb]
\centering
\includegraphics[width=0.5\columnwidth,angle=0]{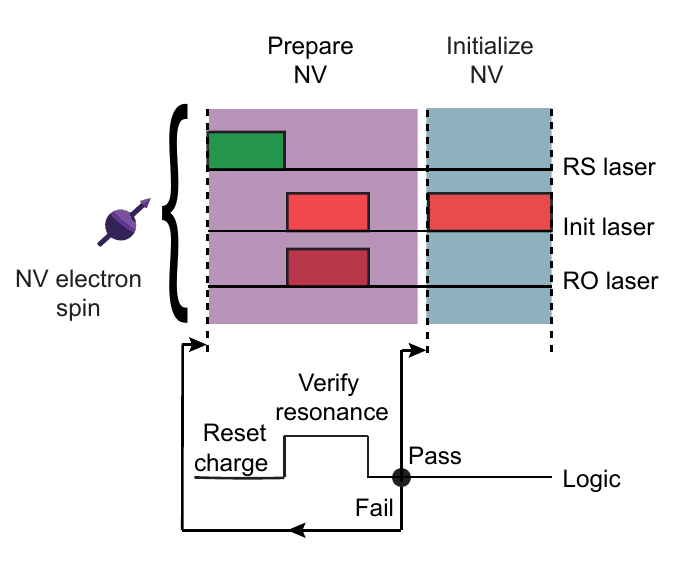}
\caption{NV preparation and initialization. \textbf{NV preparation}: The NV centre is prepared in the negative charge state and brought on resonance with the lasers used for the initialization (Init laser) and readout (RO laser) steps. We simultaneously apply the Init and RO lasers for 150 µs and count the number of detected photons (wavelength $\sim$ 637 nm, RO laser resonant with $m_s = 0 \leftrightarrow E_x$ transition and Init laser resonant with $m_s = \pm 1 \leftrightarrow E_{1,2}$ transition) \cite{Robledo_Nature2011}. If the number of detected photons exceeds a certain threshold, the NV is in the negative charge state and on resonance with both lasers, and the sequence proceeds to the next step. If not, the charge reset laser (RS, wavelength $\sim$ 515 nm) is applied for 1 ms and the same process is repeated until success \cite{Robledo_Nature2011}. \textbf{NV initialization}: the NV electron spin is initialized into the $m_s = 0$ state through spin pumping on the $E_{1,2}$ transition (Init laser, 100 µs) \cite{Robledo_Nature2011}.}
\label{fig:NV_preparation}
\end{figure}

\newpage
\subsection{Two-qubit gates}

Our native two-qubit gates are performed by applying dynamical decoupling sequences on the NV electronic spin, in resonance with a targeted nuclear spin \cite{taminiau_universal_2014,bradley2019ten}. This leads to controlled rotations which are equivalent to the standard CNOT gate up to single qubit rotations. See Supplementary Fig. \ref{fig:dd_gates} for the gate definitions.

\begin{figure}[!htb]
\centering
\includegraphics[width=0.9\columnwidth]{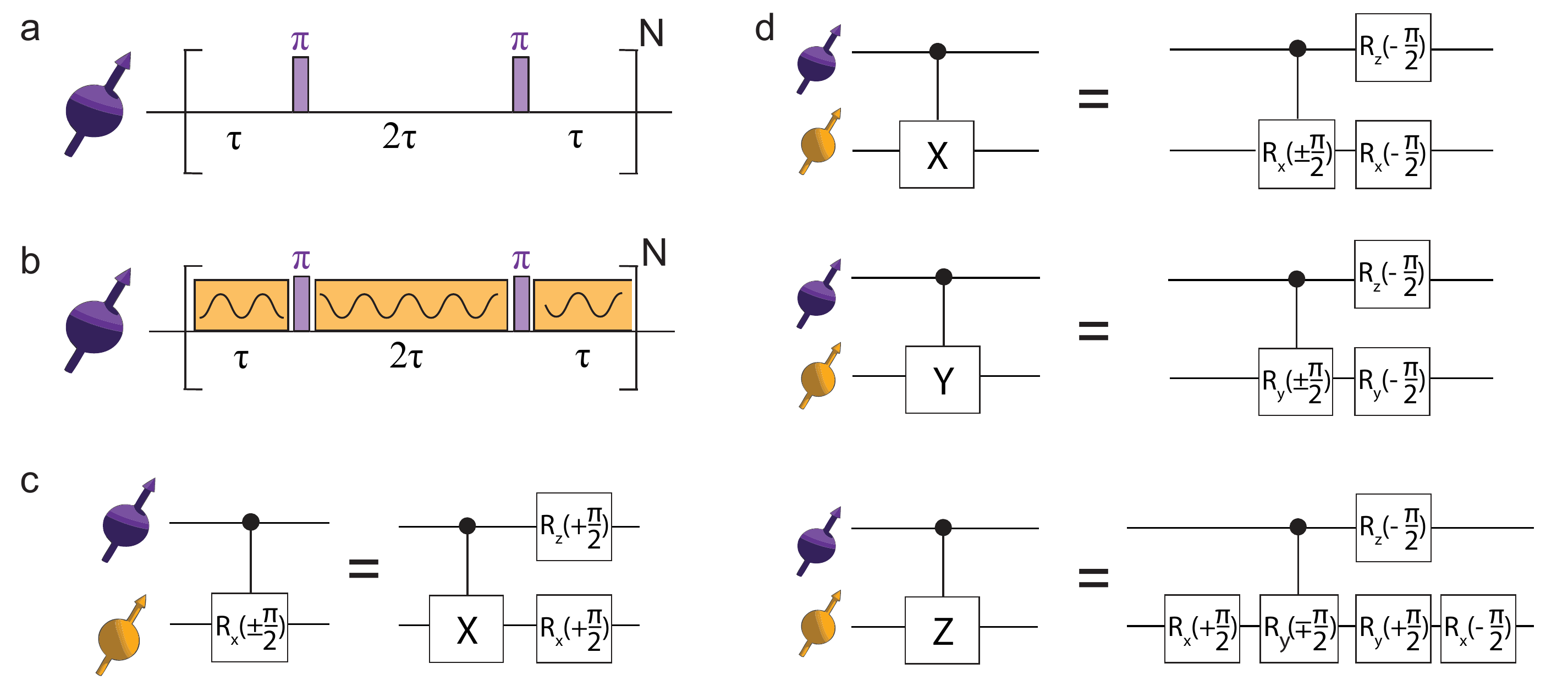}
\caption{Experimental two-qubit gates. a) Sequence for DD gate: A decoherence-protected electron-nuclear gate is realized by a dynamical decoupling sequence of $N$ equally spaced $\pi$-pulses on the electron spin of the form ($\tau_r -\pi - \tau_r)^{N}$, and setting $\tau$ to be resonant with one of the $^{13}C$ nuclear spins \cite{taminiau_universal_2014,bradley2019ten}. This design relies on the presence of a significant hyperfine-interaction component perpendicular to the applied magnetic field \cite{taminiau_universal_2014}. b) Sequence for DDRF gate: if the perpendicular hyperfine coupling is small, we perform two-qubit gates by interleaving the dynamical decoupling sequence with resonant radio-frequency pulses \cite{bradley2019ten}. c) Both sequences realize an electron-controlled nuclear gate, ${\rm \it{controlled}}-R_{x}$($\pm \pi/2$), which is equivalent to a standard CNOT gate up to single qubit rotations. Note that ${\rm \it{controlled}}-R_{x}(\pm \pi/2) = \ket{0}\bra{0} \otimes R_{x}(+\pi/2) +  \ket{1}\bra{1} \otimes R_{x}(-\pi/2)$. d) Implementation of the standard controlled gates using our two-qubit gate schemes and extra single qubit rotations. ${\rm \it{controlled}}-R_y(\pm \pi/2)$ can be directly implemented using $R_x(\pm \pi/2)$ by adding phase-shift gates. }
\label{fig:dd_gates}
\end{figure}

\newpage
\subsection{Initialization and final readout of the data qubits}

\begin{figure}[!htp]
\centering
\includegraphics[width=1\columnwidth]{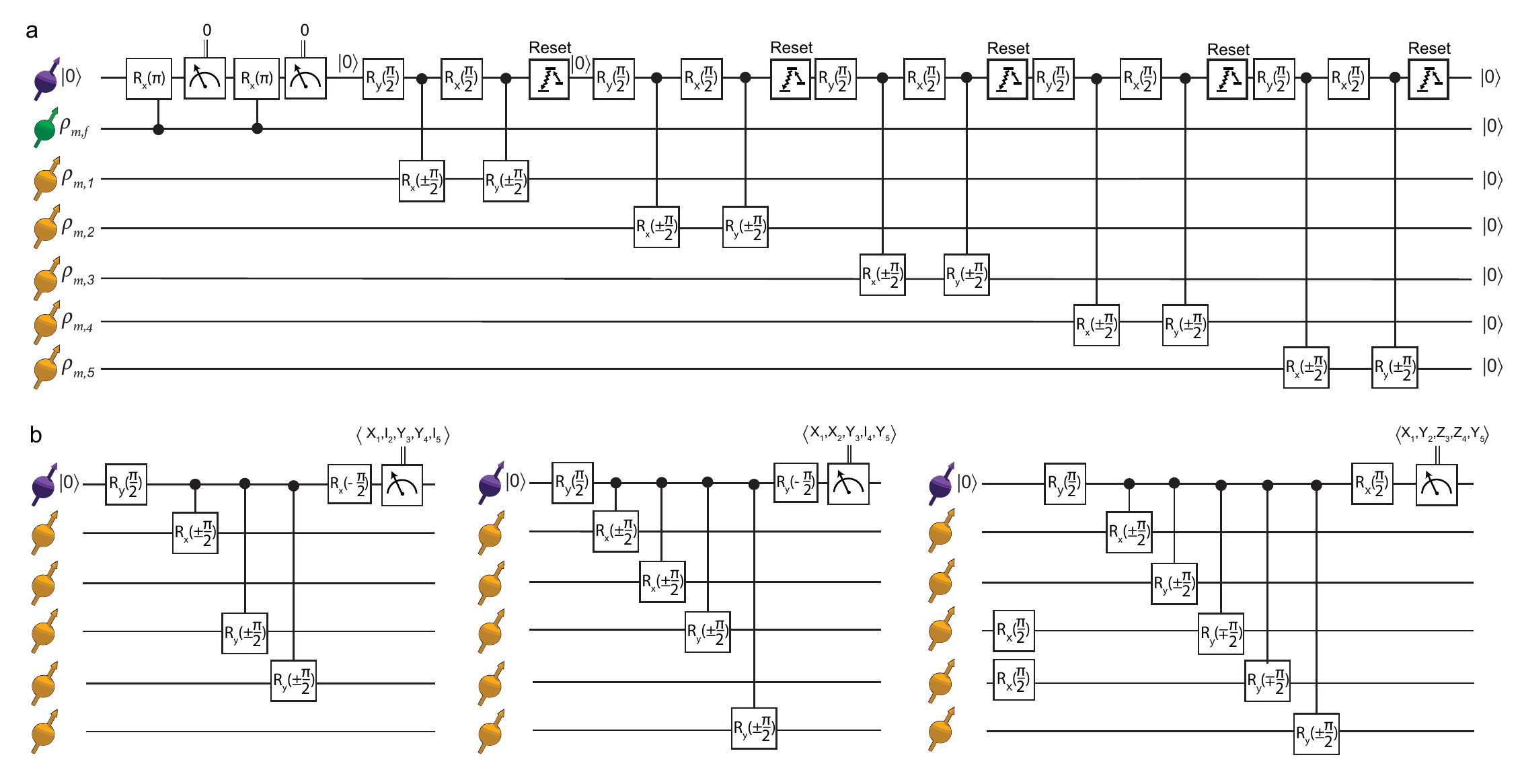}
\caption{Initialization and readout of the data qubits. a) Initialization of the nuclear-spin qubits. First, the $^{14}$N nuclear-spin flag qubit is initialized using measurement-based initialization sequence (performed twice to improve the initialization fidelity) \cite{bradley2019ten}. Afterwards, the five data qubits are initialized sequentially by SWAP sequences that transfer the electron spin state ($\ket{0}$) onto the data qubits. An optical pulse is used to reset the electron state to $\ket{0}$ following each SWAP step. b) Measuring multi-qubit operators of the data qubits is performed by mapping the required correlation to the ancilla qubit (through controlled qubit rotations) and then reading out the ancilla qubit. As each additional electron-controlled nuclear gate adds a $\pi/2$ phase shift to the electron spin (Supplementary Fig. \ref{fig:dd_gates}), the final phase of the $\pi/2$ readout pulse accounts for the number of the spins being measured (i.e., the phase depends on the number of non-identity terms in the measured operator). Three examples are shown that capture the main idea. These type of measurements are used both to realize non-destructive stabilizer measurements, as well as to tomographically measure the data qubits to characterize the final state.}
\label{fig:init_tomo}
\end{figure}

\newpage
\subsection{Circuit compilation}

\begin{figure}[!htb]
\centering
\includegraphics[width=0.85\columnwidth,angle=0]{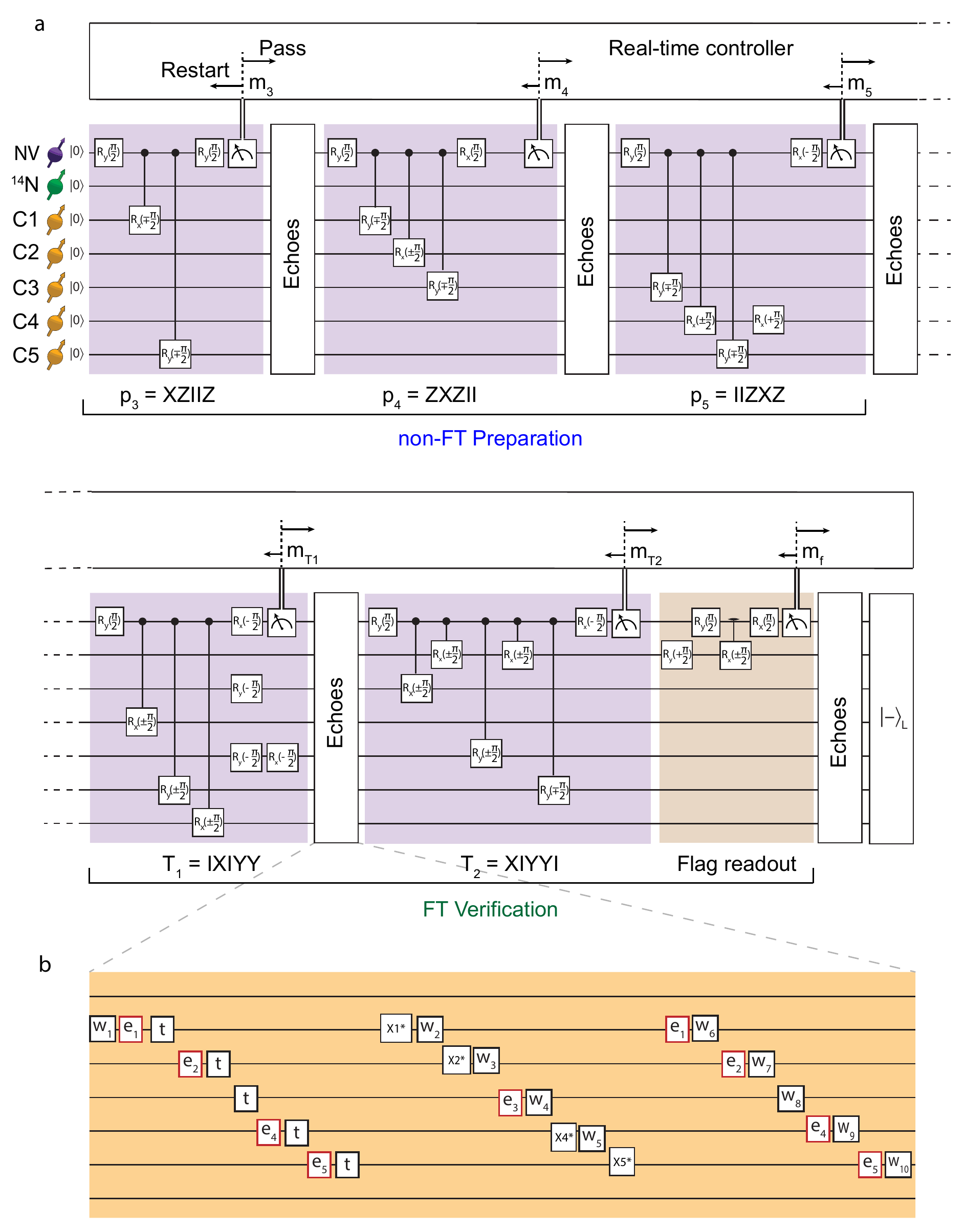}
\caption{Experimental implementation of the logical state encoding scheme. The circuit diagram in Fig.~3a in the main text is converted to our native gates and then compiled to minimize the number of single qubit gates. The $^{13}$C nuclear spins are first initialized into $\ket{00000}$ (Supplementary Fig.~\ref{fig:init_tomo}).
b) We apply echo stages on the nuclear spins between the stabilizer measurements to mitigate the nuclear-spin decoherence and unwanted couplings. The echos are applied in asynchronous form: for qubit C3 only a single echo pulse is applied per stage, so that it is also decoupled from the other 4 qubits on which we apply two echo pulses per stage. This simultaneously enhances the nuclear-spin coherence time and avoids effects of nuclear-nuclear interactions between the nuclear-spin qubits. Boxes $e_1, e_2,...,e_5$ correspond to spin echo pulses on those spins, with unique lengths calibrated to maximize the pulse fidelity. Boxes $X_1^*,..,X_5^*$ are rephasing points, and at these points single-qubit rotations can be applied (in case of a required basis transformation). Boxes $W_1, W_2,...,W_{10}$ and $t$ are unique waiting duration times derived from a set of simultaneous equations following the concepts presented in Ref.~\cite{bradley2019ten}. } 
\label{fig:Encoding_gates}
\end{figure}

In order to implement the circuits presented in the main text, we first translate all gates into our native gates (Supplementary Fig.~\ref{fig:dd_gates}) and afterwards compile the circuit to reduce the total number of single-qubit gates. Supplementary Fig.~\ref{fig:Encoding_gates} shows the compiled circuit diagram for the encoding circuit presented in Fig. 3a in the main text. Note that the same approach is also taken when for performing stabilizer measurements on the logically encoded state (Fig. 5a in the main text).
\begin{table}[!htb]
  \centering
  \begin{tabular}{c|c}
  Measurement			& Success probability         \\ \hline \hline
   $p_3$	    & 0.39        	\\
   $p_4$ 	    &  0.41 \\
   $p_5$	    & 	0.41  \\
   $T_1$	    & 	0.62   \\
   $T_2$	    &  0.49   \\
   Flag	        &  0.76   \\
   \end{tabular}
  \caption{Success probabilities for the stabilizer measurements in the encoding scheme (i.e., probabilities to obtain +1 eigenvalues). }

  \label{Tab:encoding_prob}
\end{table}

\subsection{Echo sequences for the data qubits}
 
An important issue is how to mitigate decoherence of the data qubits in order to leave few errors to be dealt with by error correction. The used circuits involve up to 40 two-qubit gates (including those required for the initialization and readout of the data qubits); the typical gate time is 0.5 to 1 ms (Supplementary Table 9). This makes the total sequence time much longer than the dephasing time of the individual nuclear spins ($T_2^* = 3 - 17$ ms for the data qubits, see Supplementary Table.~\ref{Tab:SpinsCoh}). An effective way to overcome this dephasing is to use spin echoes, which can increase the coherence times to several seconds \cite{bradley2019ten}. The designed echo sequence should minimize any idle waiting time during which the electron spin is in a superposition state (i.e., not in an spin eigenstate). We found that using two echo stages provides a general solution such that the nuclear spins can refocus at the required points (see the supplementary material of Ref.~\cite{bradley2019ten} for details).

An additional challenge here is that the small couplings between the nuclear-spin data qubits become non-negligible for the long sequences implemented here (up to 100 ms when including the echo pulses) \cite{abobeih2019atomic}. From the measured coupling strengths (Supplementary Table~\ref{Tab:nuclear-nuclear couplings}), we find that the most relevant ones are those between data qubits C3$\leftrightarrow$C2 and C3$\leftrightarrow$C5 (16.90(4) Hz and 12.96(4) Hz respectively). To overcome this we use asynchronous echo sequences, i.e., we do not echo all the spins simultaneously. By applying a single echo pulse on qubit C3, it is also decoupled from the other 4 qubits, for which we apply two echo pulses (see Supplementary Fig. \ref{fig:Encoding_gates} for illustration). We adopt this approach for all echo stages used in this work.

\newpage

\section{Additional data}

\begin{figure}[!htb]
\centering
\includegraphics[width=0.4\columnwidth]{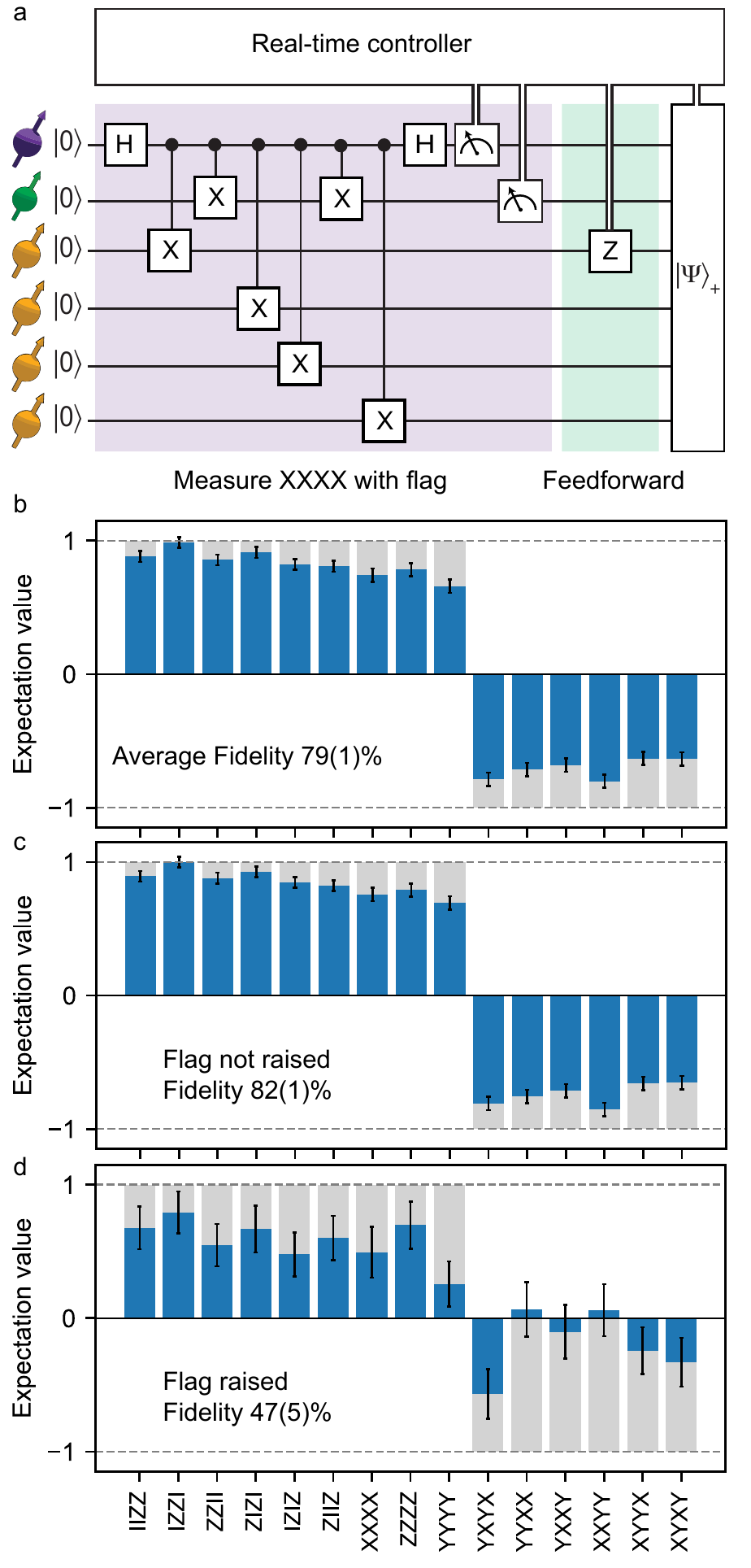}
\caption{Non-destructive stabilizer measurements with a flag and real-time feedforward. a) Circuit diagram for the deterministic preparation of a 4-qubit GHZ entangled state ($\ket{\Psi}_{+} =\ket{0000}+\ket{1111}/{\sqrt{2}}$) using a flagged measurement of the stabilizer $XXXX$. b) Measured expectation values of the 15 operators that define the ideal state. The average obtained fidelity is 0.79(1). c) Data post-selected on the flag not being raised. The obtained fidelity with the target state is 0.82(1). d) When the flag is raised, the obtained fidelity is 0.47(5). Grey bars show the ideal expectation values. Note that we perform this measurement as a test of the circuit, but that the flag information in this case does not carry any specific significance.}
\label{fig:det_ent_flag}
\end{figure}

\begin{figure}[!htb]
\centering
\includegraphics[width=\columnwidth]{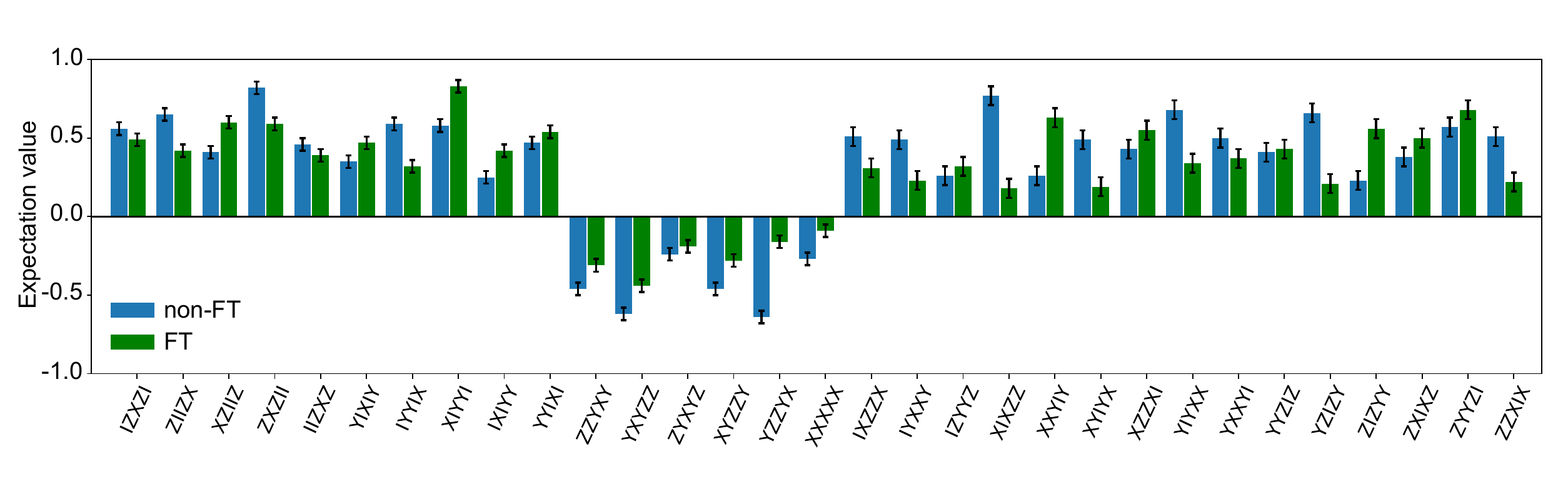}
\caption{Measured expectation values of the 31 operators that define the encoded state (for the circuit in Fig. 3 in the main text).}
\label{fig:Encoding_tomo}
\end{figure}

\clearpage
\section{Parameters of the nuclear-spin qubits}

\begin{table}[h!]
  \centering
  \begin{tabular}{c|c|c|c|c|c}
  			& $\omega_{0} / 2\pi$ [Hz]& $\omega_{-1} / 2\pi$ [Hz] & $\omega_{+1} / 2\pi$ [Hz] & $A_\parallel /2\pi$ [kHz] & $A_\perp/2\pi$ [kHz] \\ \hline \hline
   C1	    &	431956(1)       & 	469020(1)   & 396542(1)  &   -36.308(1)  &	26.62(4) 	\\
   C2	    &   431962(1)       &   408317(1)   & 457035(1)  &   24.399(1) 	&	24.81(4) 	\\
   C3	    &	431961(1)       & 	218828(1)   & 645123(1)  &   213.154(1)  &	3.0(4)	\\
   C4 	    &	431958(1)       &   413477(1)   & 454427(1)  &   20.569(1) 	& 	41.51(3) 	\\
   C5	    &	431962(1)       &   480625(1)   & 383480(40) &   -48.58(2) 	&	9(2)	\\
    $^{14}$N	    &	5069110(1)      &   2884865(1)  & 7263440(1) &   2189.288(1)   &	- 	\\
   \end{tabular}
  \caption{\textbf{Precession frequencies and hyperfine couplings for the nuclear spin qubits.} $\omega_{0}$, $\omega_{-1}$, and $\omega_{+1}$ are the measured nuclear precession frequencies for the $m_s = 0, -1$ and $+1$ electron spin projections respectively, obtained from least-squares fits of Ramsey signals. $A_\parallel$ and $A_\perp$ are the hyperfine interaction components parallel and perpendicular to the applied magnetic field. For the $^{14}$N spin the frequencies given are for the $m_I = 0 \leftrightarrow m_I = -1$ transition, and the parallel hyperfine component is taken as $(\omega_{+1} - \omega_{-1})/2$. }
  \label{Tab:Spins}
\end{table}

\begin{table}[ht!]
  \centering
  \begin{tabular}{c|c|c|c|c}
  			& $T^{*}_{2}$ ($m_{s} = -1$) [ms] & $T^{*}_{2}$ ($m_{s} = 0$) [ms] & $T_{2}$  ($m_{s} = -1$) [s] & $T^{\alpha=256}_{2}$ ($m_{s} = -1$) [s] \\ \hline \hline
   C1	    & 9.2(8)    & 9.1(5)     & 0.53(6) &  6.8(8)      \\
   C2	    & 15.6(8)   & 17.2(4)    & 0.62(3) &  12.9(4)    \\
   C3	    & 12.0(6)   & 10.0(3)    & 0.77(4) &  25(4)      \\
   C4 	    & 11.9(5)   & 12.3(3)    & 0.68(3) &  7.4(8)      \\
   C5	    & 3.7(2)    & 3.6(2)     & 0.59(2) &  13(2)    \\
   $^{14}$N	& 23.2(7)   & 25.1(7)    & 2.3(2)  &  63(2) \\
   \end{tabular}

  \caption{\textbf{Coherence times for the nuclear-spin qubits \cite{bradley2019ten}.} $T^{*}_{2}$ is the dephasing time obtained from a least-squares fit of Ramsey signals, with evolution for the $m_{s} = -1$ and $m_s = 0$ electron spin projections. $T_{2}$ is coherence time obtained from a least-squares fit of a spin echo experiment, with free evolution measured for the $m_{s} = -1$ electron spin projection. $T^{\alpha=256}_{2}$ is coherence time obtained from a least-squares fit of a dynamical decoupling experiment with $\alpha = 256$ pulses, with the electron spin in the $m_{s} = -1$ spin projection.}
  \label{Tab:SpinsCoh}
\end{table}

\begin{table}[ht!]
  \centering
  \begin{tabular}{c|c|c|c|c|c|c|c}
  			& N & $\tau$ (\textmu s)& Gate type & Gate duration (\textmu s) & $F_{gate}$ & $F_{init}$ & RF $\pi$-pulse duration (\textmu s) \\ \hline \hline
   C1	    &	44          & 	 7.218     & DD   & 635    & 0.99(1)   & 0.985(5)  & 1096	    \\
   C2	    &   32          &   6.540      & DD   & 419    & 0.99(1)   &  0.980(5) & 1606	    \\
   C3       &   48          &   16.204     & DDRF & 1556   & 0.97(1)   &  0.965(5) & 1330      \\
   C4 	    &	22          &   11.250     & DD   & 495    & 0.97(1)   &  0.970(5) & 929	    \\
   C5	    &	90          &   4.932      & DD   & 888    & 0.95(1)   &  0.985(5) & 1173	    \\
   $^{14}$N	&	12          &   16.204     & DDRF & 389    &     -     &  0.997(11) & 278       \\
   \end{tabular}

  \caption{\textbf{Gate parameters and fidelities for the nuclear spin qubits.} $N$ and $\tau$ are the parameters used for the dynamical decoupling sequence to implement electron-nuclear two-qubit gates (Supplementary Figure 3), where $N$ is the total number of electron $\pi$-pulses, and 2$\tau$ is the interpulse delay.  Gate type refers to whether the dynamical decoupling sequence is interleaved with RF pulses (DDRF) or not (DD), see Supplementary Fig.~\ref{fig:dd_gates}. $F_{init}$ is the measured initialization fidelity for the nuclear-spin qubits \cite{bradley2019ten}. $F_{gate}$ is the gate fidelity estimated from the electron-nuclear Bell-state fidelity \cite{bradley2019ten}. For the $^{14}$N flag qubit, Bell-state tomography is challenging because the optical read-out of the ancilla induces rapid dephasing. A lower bound of 0.94(1) is extracted from such measurements, but the fidelity is likely much higher.}
  \label{Tab:SpinsGate}
\end{table}

\begin{table}[h!]
  \centering
  \begin{tabular}{c|c|c|c|c|c}
  			& C1        &    C2     &   C2        &  C4     &    C5            \\ \hline \hline
   C1	    &  -        &   2.64(4) &   2.8(1)   &  2.2(4)  &	2.6(1) 	\\
   C2 	    &  2.64(4)  &   -       &	16.90(4)   &  1.08    & 	1.42 	\\
   C3	    & 	2.8(1)  &   16.90(4) &	-         &  3.12(8) &	12.96(4)	\\
   C4	    & 	2.2(4)  &   1.08    &	3.12(8)    &  -      &	4.0(1)	\\
   C5	    &  2.6(1)   &   1.42    &	12.96(4)   &  4.0(1) &	-	\\
   \end{tabular}
  \caption{\textbf{Qubit-qubit coupling.} Measured coupling strength ($C_{ZZ}$) in Hz between the nuclear-spin data qubits. Values without uncertainties are calculated from the nuclear spin coordinates, rather than measured directly \cite{abobeih2019atomic}.}
  \label{Tab:nuclear-nuclear couplings}
\end{table}

\begin{figure}
\includegraphics[width=0.8\linewidth]{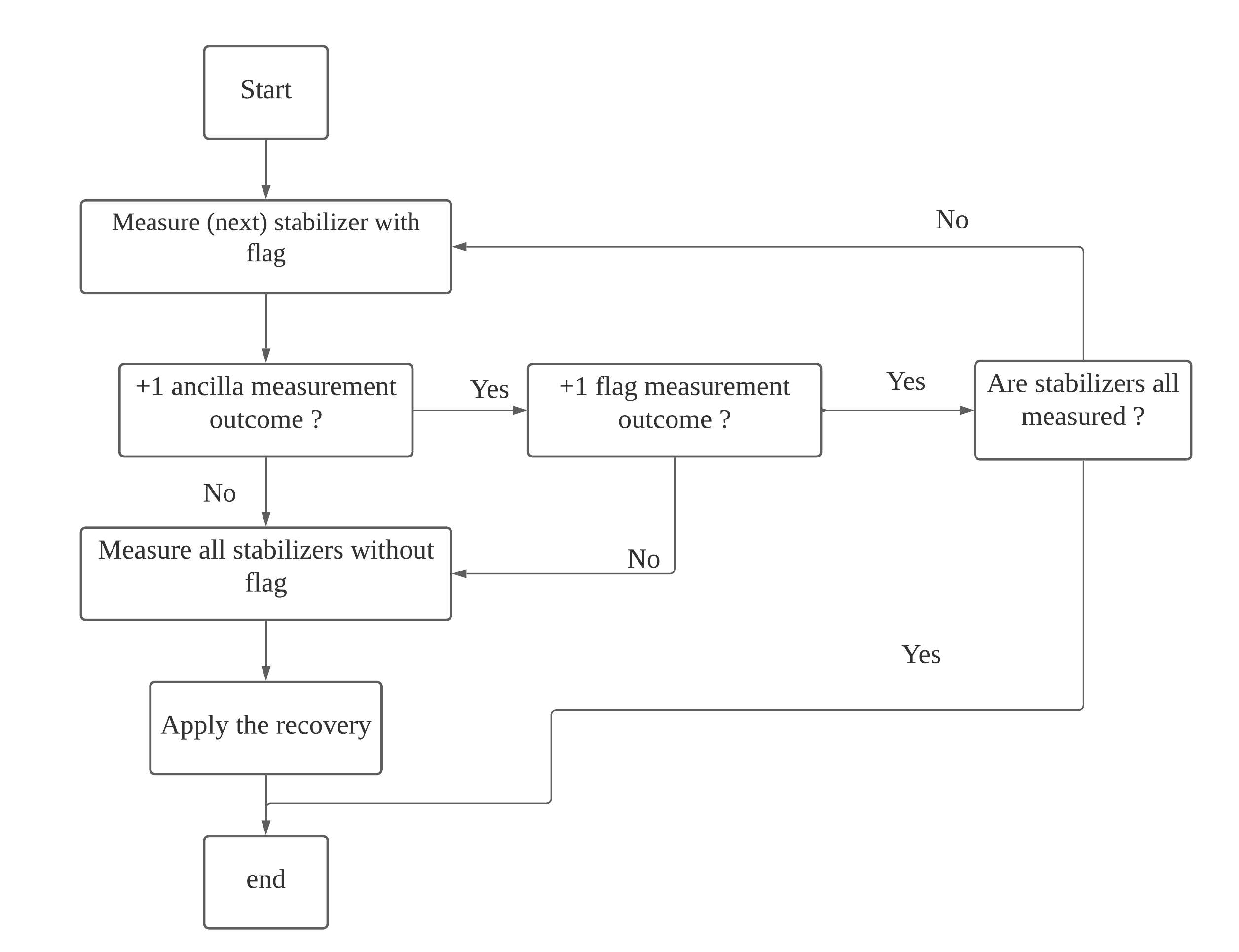}
\caption{The flowchart of a full flag error correction cycle proposed by Chao and Reichardt \cite{chao2018}. For the 5-qubit code, the stabilizers are $s_1 = XXYIY, s_2 = YXXYI, s_3 = IYXXY$ and  $s_4 = YIYXX$. The stabilizer measurements without flag give a 4-bit error syndrome, which determines the recovery operation. Note that the error syndrome is interpreted differently depending on whether there is a raised flag. In the experiments presented in this work, we apply one stabilizer measurement of this error correction cycle and characterize the resulting post-measurement state. 
}
\label{fig:flow_chart_flag}
\end{figure}

\begin{figure}
\centering
\includegraphics[width=0.55\linewidth]{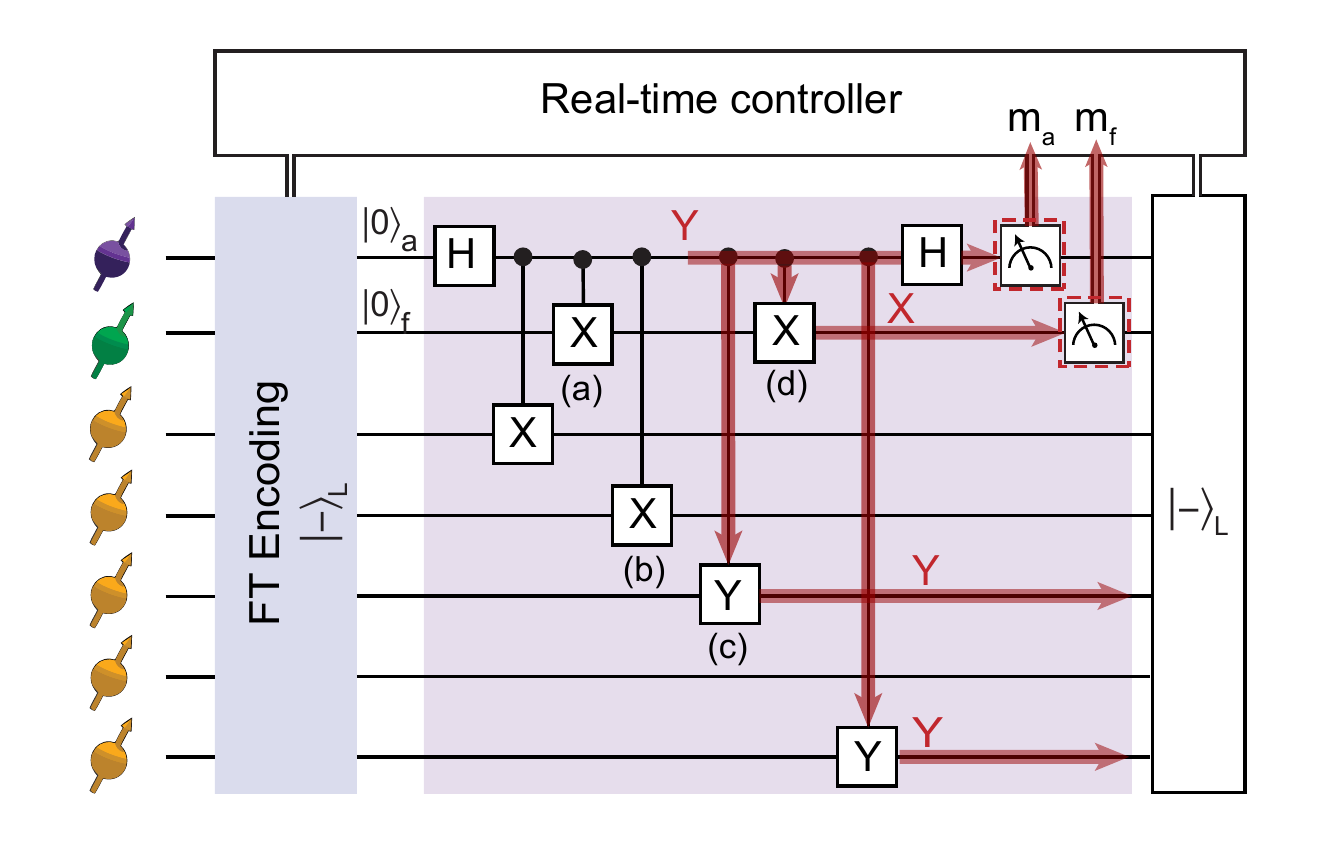}
\caption{The circuit for measuring the stabilizer $s_1 = XXYIY$ with flag on the encoded state. Given this specific order of gates, the $Y$ error on the ancilla qubit is a malicious fault, as it will propagate to the data qubits to become $Y_3X_5 = -Z_4 \cdot \overline{Z}\cdot  s_1 \cdot s_3 \cdot s_4$. This 2-qubit error $Y_3Y_5$ will lead to a logical $Z$ error after a round of noiseless error correction. However, this $Y$ error will also propagate to the flag ancilla qubit and flip its measurement outcome, which provides additional information: a flipped flag measurement heralds the possibility of such a malicious fault.
}
\label{fig:flagged_XXYIY}
\end{figure}

\begin{table}
	\centering
	\begin{tabular}{|c|c|c|c|}
	    \hline
		gate $(b)$:\, fault &  data qubit error  & gate $(c)$:\,fault & data qubit error\\
		\hline
		XI (YI) & $Y_3Y_5$  & XI (YI) & $Y_5$\\
		XX (YX) & $X_2Y_3Y_5 \equiv X_1$  &XX (YX) & $X_3Y_5$\\
		XY (YY) & $Y_2Y_3Y_5 \equiv X_1Z_2$ &XY (YY) & $Y_3Y_5$\\
		XZ (YZ) & $Z_2Y_3Y_5 \equiv X_1Y_2$  & XZ (YZ) & $Z_3Y_5$\\
		\hline
		&&&\\
		\hline
		gate $(a)$:\, fault &  data qubit error & gate $(d)$:\,fault & data qubit error\\
		\hline
		XZ (YZ) & $X_2Y_3Y_5 \equiv X_1$  & XX (XY) & $Y_5$\\
		XI (YI) & $X_2Y_3Y_5 \equiv X_1$  &YX (YY) & $Y_5$\\
		IX (IY) & $I$  &IX (IY) &  $I$\\
		ZX (ZY) & $I$  &ZX (ZY) &  $I$\\
		\hline 
		
	\end{tabular}
\caption{Single faults that can flip the flag qubit measurement in the stabilizer $s_1 = XXYIY$ measurement circuit (see Supplementary Fig.~\ref{fig:flagged_XXYIY}). Such a fault is either induced by a faulty two-qubit gate or a single-qubit idling error after the gate.
A readout error of the flag qubit can lead a different correction but does not induce an error on the output state directly.
Note that here we assume only one fault during the whole circuit including the state preparation. These 2-qubit Pauli errors are arranged in the order of control and target qubit respectively and they result in certain data qubit errors on the output state.
The errors in parentheses lead to the same data qubit errors. Syndromes of these resulting data qubit errors remain distinguishable, hence a perfect round of stabilizer measurements would identify and correct them (Supplementary Table \ref{tab:distinguishable_syndrome}).
}
\label{tab:single-fault-nontrivial}
\end{table}

\section{Flag fault-tolerant quantum error correction}

In our experiment we realize fault-tolerant logical state preparation and flagged stabilizer measurements of a distance-3 5-qubit quantum error correction code. Here we first briefly review the idea of (flag) fault tolerance and distance-3 codes \cite{chao2018,chamberland2018flag}. Then we analyze the error syndromes in context of the experimentally applied stabilizer measurement (Supplementary Fig.~\ref{fig:flagged_XXYIY}). In the next section \ref{proof:ft} we then provide a proof of the fault-tolerance of our heralded preparation protocol (Fig.~3a in the main text).

A distance-3 code can correct an arbitrary single-qubit error. This ability is, however, based on the assumption of syndrome extraction by ideal stabilizer measurements. In practice, one has  to design an error syndrome extraction protocol which works when the operations involved are noisy. For distance-3 codes, a unit of fault-tolerant quantum error correction or syndrome extraction satisfies the following criteria \cite{aliferis2005quantum,chamberland2018flag}: (1) If the logical input state has a single qubit error and the syndrome extraction unit is noise-free, the output state is the logical input state with its error removed; (2) If the logical input state is noise-free and there is a single fault in the syndrome extraction unit, the output state has at most a single-qubit error on top of the given input state; (3) If the input state is an arbitrary state, and there is at most one single fault in the syndrome extraction unit, then the output state is any encoded state with at most a single-qubit error. For the 5-qubit code in this work, the last requirement is automatically fulfilled. For details about fault-tolerance, we refer the reader to Refs.~\cite{aliferis2005quantum,chamberland2018flag}.

Note that a single fault constitutes the insertion of any Pauli error at a location in a circuit, where locations can be: a qubit idling step, a qubit measurement, a qubit preparation, a single or two-qubit gate. In case the location is a two-qubit gate, a single fault is the insertion of any of the 15 two-qubit Pauli errors after the action of the gate. For a single-qubit measurement a fault is getting the wrong measurement outcome.

The basic idea of flag fault-tolerance is to use the additional information provided by the flag qubit measurements to correct certain two-qubit errors \cite{chao2018,chamberland2018flag,chao2020PRXQ}. For example, we consider the measurement of $s_1 = XXYIY$ using the circuit in Supplementary Fig.~\ref{fig:flagged_XXYIY}. We assume that the logical preparation is noise-free and there is a single fault in measuring $s_1$. It can then be shown that: (1) if the flag qubit is not flipped, there is at most a single-qubit error on the output state; (2) otherwise, errors on the data qubits can only be $X_1$, $X_1Z_2$, $X_3Y_5$, $Y_3Y_5$, $Z_3Y_5$, $X_1Y_2$ and $Y_5$, 
see Supplementary Tab~\ref{tab:single-fault-nontrivial}. If we were to apply a round of noiseless error correction on the output state, these errors can be identified and corrected as they have distinguishable syndromes. Each syndrome uniquely points to what error has happened (Supplementary Table~\ref{tab:distinguishable_syndrome}).

To implement a complete flag-QEC cycle, one repeatedly measures stabilizers (with and without flag) in a conditional form, until the error on the logical qubit can be unambiguously identified \cite{chao2018, chamberland2018flag}. We summarize the protocol proposed by Chao and Reichardt \cite{chao2018} as a flow chart in Supplementary Fig.~\ref{fig:flow_chart_flag}. 

\subsection{Stabilizer measurement with flag}
In the experiments we first prepare the logical state $\ket{-}_L$ and then implement the measurement of the stabilizer $s_1= XXYIY$ using a flag qubit (Supplementary Fig.~\ref{fig:flagged_XXYIY}). This measurement is a primitive of the Chao and Reichardt protocol described above. To calculate the logical state fidelity, 
we imagine a round of perfect stabilizer measurements on the experimentally-obtained output state. This round of stabilizer measurements can use the flag information to apply optimal recoveries. 

If the flag qubit measurement is flipped, all possible errors on the output state due to a single fault are listed in Supplementary Table~\ref{tab:single-fault-nontrivial}. 
Note that these errors are obtained with the assumption that the input state is a logical state without error and there is only one fault in measuring $s_1$. However, errors outside the table could occur due to an incoming error on the logical state or possibly multiple faults in the circuit of measuring $s_1$. In such cases, the strategy of the imagined noiseless error correction round is to simply apply the corresponding single-qubit recovery, as if the flag qubit is not flipped.
Therefore, when the flipped flag qubit measurement is taken into account, the set of correctable errors is
\begin{equation}
    \mathcal{E}' = \{I, X_1, X_3Y_5, Z_1, X_2, Y_2, Z_3Y_5, X_1Y_2, Y_3, Z_3, X_4, Y_4, Y_3Y_5, X_5, Y_5, X_1Z_2 \}.
\end{equation}
The corresponding syndromes are listed in Supplementary Table~\ref{tab:distinguishable_syndrome}.
\begin{table}[htb]
	\centering
	\begin{tabular}{|c|c|c|c|c|c|}
		\hline
		$s_1$ &$s_2$ & $s_3$ & $s_4$ & $\mathcal{E}$ & $\mathcal{E}'$\\
		\hline
		+1& +1 & +1 & +1 & $I$  & $I$ \\
		\hline
		+1& -1 & +1 & -1 & $X_1$  & $X_1$ \\
		\hline
		-1& -1 & +1 & -1 & $Z_1$  & $Z_1$ \\
		\hline
		-1& +1 & +1 & +1 & $Y_1$  & $X_3Y_5$ \\
		\hline
		+1& +1 & -1 & +1 & $X_2$  & $X_2$ \\
		\hline
		-1& -1 & -1 & +1 & $Z_2$  & $Z_3Y_5$ \\
		\hline
		-1& -1 & +1 & +1 & $Y_2$  & $Y_2$ \\
		\hline
		-1& +1 & +1 & -1 & $X_3$  & $X_1Y_2$ \\
		\hline
		-1& -1 & -1 & -1 & $Z_3$  & $Z_3$ \\
		\hline
		+1& -1 & -1 & +1 & $Y_3$  & $Y_3$ \\
		\hline
		+1& -1 & +1 & +1 & $X_4$  & $X_4$ \\
		\hline
		+1& -1 & -1 & -1 & $Z_4$  & $Y_3Y_5$ \\
		\hline
		+1& +1 & -1 & -1 & $Y_4$  & $Y_4$ \\
		\hline
		-1& +1 & -1 & +1 & $X_5$  & $X_5$ \\
		\hline
		+1& +1 & +1 & -1 & $Y_5$  & $Y_5$ \\
		\hline
		-1& +1 & -1 & -1 & $Z_5$  & $X_1Z_2$ \\
		\hline
	\end{tabular}
\caption{Each 4-bit syndrome (eigenvalues of the stabilizers $s_1=XXYIY, s_2 = YXXYI, s_3 = IYXXY$ and $s_4 = YIYXX$) can correspond to different Pauli errors. When $s_1 = XXYIY$ is measured using the circuit in Supplementary Fig.~\ref{fig:flagged_XXYIY} and the flag is raised, $\mathcal{E}'$ represents the set of correctable errors with the flag information taken into account. When the flag information is not taken into account, $\mathcal{E}$ is the set of correctable errors including all single-qubit Pauli errors. 
}
\label{tab:distinguishable_syndrome}
\end{table}

\section{Proof of fault-tolerance of the preparation scheme} 
\label{proof:ft}
In this section we provide the theoretical proof that a single fault in the encoding circuit (Fig. 3a in the main text) leads to the correct preparation of $\ket{-}_L$ plus at most a single-qubit error, assuming that the conditions for accepting the state are fulfilled. The conditions are (1) the measurement outcomes of $T_1$ and $T_2$ are compatible with the measurement outcomes $m_i$ of the logical operators $p_i$, i.e.  $m_{T_1} = m_2\times m_4\times m_5$ and $m_{T_2} = m_1\times m_3\times m_5$; (2) the flag is not raised. Otherwise the state is rejected. 

The modification of the preparation scheme with additional single-qubit gates does not change the fault-tolerance argument. Therefore, the same derivation applies to preparing other basis states by adding and then compiling transversal logical gates, as well as when we execute echo pulses or use different gate decompositions. In the experimental realization we additionally condition on the preparation runs which give $m_3=+1, m_4=+1, m_5=+1$ since these measurement outcomes ($m_s=0$ NV electron spin state) are more reliable. Such heralding does not affect the fault-tolerance arguments and we provide the more general proof here. In this section, we denote controlled-NOT and controlled-Y gates as $CX$ and $CY$.

First, we note that for this 5-qubit code, any state $\ket{-}_L$ with more than 1 Pauli error is equivalent to either a $\ket{+}_L$ with at most 1 Pauli error or a $\ket{-}_L$ state with at most 1 Pauli error. This is due to the code being `perfect': the states $\ket{\pm}_L$ plus any single-qubit error are all orthogonal and there are $2\times (1+15)=2^5=32$ such states, spanning the full 5-qubit Hilbert space. 

Hence we just need to prove that the preparation circuit does not lead to the state $\ket{+}_L$ with at most 1 Pauli error. For this it is useful to tabulate some incarnations of the logical $Z$ and logical $Y$ as these can bring $\ket{-}_L$ to $\ket{+}_L$ up to a global phase, see Supplementary Table~\ref{tab:logicals}.

\begin{table}[htb]
	\centering
	\begin{tabular}{c|c|c|c}
		$Z_L \equiv$ &  $Z_1 Z_2 Z_3 Z_4 Z_5$ & $-Y_3Z_4Y_5$ (cyclic perm.) &  $-X_2 X_3 Z_5$ (cyclic perm.)\\
		$Y_L \equiv$ & $Y_1 Y_2 Y_3 Y_4 Y_5$ & $-Z_2 Z_3 Y_5$ (cyclic perm.) &  $-X_1 X_4 Y_5$ (cyclic perm.) \\
		$X_L \equiv$ & $X_1 X_2 X_3 X_4 X_5$ & $-X_2 Y_4 Y_5$ (cyclic perm.) & $-Z_1 Z_4 X_5$ (cyclic perm.)
			\end{tabular}
\caption{Incarnations of the logical operators which can be obtained by multiplying the logical operators by stabilizers (weight-3 incarnations have minus signs). Any cyclic permutation (e.g., $-X_2X_3Z_5 \rightarrow -Z_1X_3X_4  \rightarrow -Z_2X_4X_5$) of a logical operator is an equivalent logical operator.}
\label{tab:logicals}
\end{table}

Throughout this proof, we assume Pauli frame corrections by keeping track of detected errors in classical logic. The classical information changes our interpretation of the final measurement outcome, which can be basically interpreted as applying noiseless recovery according to the measurement outcomes (Supplementary Table~\ref{tab:Pauli_frame_correction}).

\begin{table}[htb]
	\centering
	\begin{tabular}{|c|c|c|c|}
		\hline
		$m_3$ & $m_4$ & $m_5$ & correction \\
		\hline
		+1 & +1 & +1 & $I$ \\
		\hline
		+1 & +1 & -1 & $Z_4$ \\
		\hline
		+1 & -1 & +1 & $Z_2$ \\
		\hline
		+1 & -1 & -1 & $X_3$ \\
		\hline
		-1 & +1 & +1 & $Z_1$ \\
		\hline
		-1 & +1 & -1 & $X_5$ \\
		\hline
		-1 & -1 & +1 & $Z_1Z_2$ \\
		\hline
		-1 & -1 & -1 & $Z_1X_3$ \\
		\hline
	\end{tabular}
\caption{Pauli corrections to bring the prepared state to $\ket{-}_L$ (Fig.~3a in the main text). The correction anti-commutes (resp. commutes) with all $p_i$ with $m_i=-1$ (resp. $m_i=+1$) and commutes with $p_1$ and $p_2$. The corrections in the look-up table are not unique as one can apply a stabilizer or any incarnation of a logical $X$ to them which leaves the state $\ket{-}_L$ unchanged (modulo overall phase).}
\label{tab:Pauli_frame_correction}
\end{table}

Looking at the circuit in Fig.~3a in the main text, we can assume that the fault appears either in the non-FT preparation circuit and the verification circuit is fault-free, or vice versa. So we consider these cases separately as follows.
\\

\noindent \textbf{Case A: if the only fault occurs in the verification circuit} 

\noindent Since the only fault occurs in the verification circuit, the preparation circuit prepares the state $\ket{-}_L$ with no errors. We prove that the circuit is fault-tolerant by considering different cases:

\begin{itemize}
\item Single-qubit faults on data qubits (in idling or after gates) during the verification. These have the effect of either leading to a single-qubit error on the output state, or leading to a measurement of $T_1$ or/and $T_2$ which does not satisfy the consistency check (in which case the output state is not accepted).

\item A single measurement fault in the measurement of $T_1$, $T_2$ or the flag ancilla measurement leads to the state not being accepted.

\item A two-qubit fault after one of the two-qubit gates in $T_1$. If the action of this fault on the ancilla qubit is $Z$ or $Y$, then it flips the ancilla and the state is not accepted. Hence we assume the action on the ancilla qubit is $X$. If the error $XP \equiv X \otimes P$, where $P$ is some Pauli $I, X,Y, Z$ acting on a data qubit,
occurs after the first $CX$, it leads to outgoing error $P_2Y_4Y_5$ which is equivalent to $(X_2P_2)X_L$, hence a single-qubit outgoing error. If $XP$ occurs after $CY$ in the middle, it leads to outgoing error $P_4Y_5$: (1) $Y_4Y_5$ is equivalent to $X_2 X_L$, (2) $X_4Y_5$ and $Z_4Y_5$ lead to the $T_2$ outcome being flipped and no acceptance, (3) $P=I$ corresponds to the single-qubit error $Y_5$. If $XP$ occurs after the (last) $CY$, it leads to a single-qubit outgoing error $P_5$.

\item A two-qubit fault after one of the two-qubit gates in $T_2$. If the action of this fault on the ancilla qubit is $Z$ or $Y$, then it flips the ancilla qubit and the state is not accepted. Hence we assume that the action on the ancilla qubit is $X$. 

Consider first the two-qubit gates between ancilla and data qubits. If the fault $XP$ occurs after the $CX$, then it does not flip the flag qubit, and leads to outgoing error $P_1 Y_3 Y_4$ which is equivalent to $(X_1 P_1) X_L$, hence inducing a single-qubit error.
If the fault $XP$ occurs after $CY$ in the middle, it flips the flag qubit and the state is not accepted. 
If the fault $XP$ occurs after the second $CY$, it leads to a single qubit error $P_4$. 

Now consider $CX$ gates between ancilla and flag ancilla qubit. Note that an error on the flag qubit cannot propagate to the data qubits, we thus only consider a Pauli $X$ error on the ancilla qubit. An $X$ error on ancilla after the first $CX$ leads to $Y_3Y_4$, which is equivalent to $X_1X_L$. An $X$ error on the ancilla after the second $CX$ leads to the single-qubit error $Y_4$. 
\end{itemize}

\begin{table}[htb]
	\centering
	\begin{tabular}{c|c|c|c|c|c|c|c|}
		 &  $p_1$ & $p_2$ &  $p_3$ & $p_4$ & $p_5$ & $T_1=p_2 p_4 p_5$ & $T_2=p_1 p_3 p_5$ \\
		 \hline
		 $Y_1$ &    & F & F & F &    &     & F  \\		 \hline
		 $X_1$ &    & F &    & F &    &     &     \\		 \hline
		 $Z_1$ &    &    & F &    &     &  &F  \\		 \hline
		 $Y_2$ & F &    & F & F &    & F   &   \\		 \hline
		 $X_2$ & F &    & F &    &    &      &   \\		 \hline
		 $Z_2$ &    &    &     & F &    & F  &   \\		 \hline
		 $Y_3$ & F &    &     & F & F&     &   \\		 \hline
		 $X_3$ &    &    &     & F & F&     & F  \\		 \hline
		 $Z_3$ & F &    &     &    &   &     & F  \\		 \hline
		 $Y_4$ & F & F &    &     & F &   &   \\		 \hline
		 $\textcolor{red}{X_4}$ & F & F &    &    &     & F & F  \\		 \hline
		 $\textcolor{red}{Z_4}$ &    &     &    &    & F & F  & F  \\		 \hline
		 $Y_5$ &    & F & F &    & F &    &    \\		 \hline
		 $X_5$ &    &    & F &    & F &   F &   \\		 \hline
		 $Z_5$ &    & F &    &    &    &  F  &   \\		 \hline
			\end{tabular}
\caption{Effect of single qubit errors. $\mathrm{F}$ denotes that the eigenvalue is flipped upon application of the error, i.e., the error anti-commutes with the operator. For the logical state $\ket{+}_L$, both measurements of the verification checks $T_1$ and $T_2$ should give $-1$, which fails the verification. However, $\ket{+}_L$ with the errors highlighted in red ($X_4$ and $Z_4$) would pass the verification test as both verification checks are flipped.}
\label{tab:flips}
\end{table}


\noindent\textbf{Case B: if the only fault occurs in the preparation circuit} 

\noindent  When the non-FT circuit gives the output $\ket{-}_L$ plus at most a single-qubit error, no matter whether this state passes the verification or not, the final output state can have at most a single-qubit error as the verification circuit is perfect. For output states $\ket{+}_L$ with at most a single-qubit error, we need to
argue: (1) either they are caught by the verification circuit (2) or they cannot occur due to a single fault in the non-FT preparation circuit.

Note that the output $\ket{+}_L$ is a $-1$ eigenstate of $p_1, p_2, p_3, p_4$ and $p_5$. If the output state of the non-FT circuit is $\ket{+}_L$ with no error, both $T_1$ and $T_2$ should give measurement outcome $-1$ as they are products of three $p_i$s. Such a state would be rejected by the verification circuit. Now we examine what possible single-qubit errors on $\ket{+}_L$ would lead to passing the verification test. We can see that these are the errors $X_4$ and $Z_4$ on top of $\ket{+}_L$, i.e., both anti-commute with the verification checks (Supplementary Table~\ref{tab:flips}).

One can fully characterize these possible bad states $X_4 \ket{+}_L$ and $Z_4 \ket{+}_L$ by their syndromes $M_i=\pm 1$ (eigenvalues of $p_1, \ldots, p_5$ if we were to measure these logical operators noiselessly). Syndromes of the possible bad states (denoted by capital letters) are given as follows
\begin{align}
\label{eq:syndrome1}
    X_4 \ket{+}_L\colon & M_3 = M_4 = M_5 = -1, \quad &M_1 = M_2 = +1,  \\
    Z_4 \ket{+}_L\colon  & M_1 = M_2 = M_3 = M_4 = -1, \quad &M_5 = +1
    \label{eq:syndrome2}
\end{align}
Now we argue that no single fault in the preparation circuit can lead to $X_4 \ket{+}_L$ or $Z_4 \ket{+}_L$. We look at various subcases:
\paragraph*{1) If the only fault occurs before measuring $p_4$ and $p_5$.} Then the Pauli corrections correctly fix the eigenvalues of $p_4$ and $p_5$ on the output so that $M_4=M_5=+1$. Both syndromes in Eq.~(\ref{eq:syndrome1}) and Eq.~(\ref{eq:syndrome2}) are therefore excluded.

\paragraph*{2) If the only fault occurs in measuring $p_4$.} Similarly, $M_5$ is fixed to be $+1$ due to correct Pauli corrections, so that the syndrome in Eq.~(\ref{eq:syndrome1}) is not possible. Then for the syndrome in Eq.~(\ref{eq:syndrome2}) we argue as follows:
\begin{itemize}
    \item In the circuit of measuring $p_4$, only faults  after $CX$ in the middle can lead to non-trivial 2-qubit errors $Z_2Z_3$ and $Y_2Z_3$. These errors commute with $p_2 = ZIIZX$, hence $M_2$ is fixed to be $+1$. The syndrome in Eq.~(\ref{eq:syndrome2}) is therefore not possible. 
    \item Up to a stabilizer or a logical $X$ operator, other single faults can only induce single-qubit errors on the data qubits. Because any single-qubit error cannot anti-commute with $p_1 = IZXZI$, $p_2 = ZIIZX$ and $p_3 = XZIIZ$ at the same time,  $M_1$, $M_2$ and $M_3$ cannot be $-1$ at the same time. The syndrome in Eq.~(\ref{eq:syndrome2}) is therefore not possible.
\end{itemize}

\paragraph*{3) If the only fault occurs in measuring $p_5$.} 
\begin{itemize}
\item In the circuit of measuring $p_5$, only faults  after  $CX$ in the middle can lead to non-trivial 2-qubit errors $Z_4Z_5$ and $Y_4Z_5$. These errors both commute with $p_3 = XZIIZ$ and $p_4 = ZXZII$, i.e., both $M_3$ and $M_4$ are fixed to be $+1$. The syndromes in Eq.~(\ref{eq:syndrome1}) and Eq.~(\ref{eq:syndrome2}) are thus not possible. 

\item If the only fault occurs on an idling location of the data qubits C1 or C2, the measurement of $p_5$ is then correct. Because the induced single-qubit errors commute with $p_5 = IIZXZ$, which means the Pauli correction fixes $M_5$ to be $+1$. The syndrome in Eq.~(\ref{eq:syndrome1}) are therefore not possible. In addition, any single-qubit error on C1 or C2 cannot anti-commute with $p_1 = IZXZI$, $p_2 = ZIIZX$, $p_3 = XZIIZ$ and $p_4 = ZXZII$ at the same time, the syndrome in Eq.~(\ref{eq:syndrome2}) is also excluded.

\item Up to a stabilizer or a logical $X$ operator, other single faults can only induce a single-qubit error on the data qubits C3, C4 or C5. Such single-qubit errors cannot anti-commute with $p_3 = XZIIZ$ and $p_4 = ZXZII$ at the same time, the syndromes in Eq.~(\ref{eq:syndrome1}) and Eq.~(\ref{eq:syndrome2}) are therefore not possible. 
\end{itemize}

\bibliographystyle{naturemag}
\bibliography{bibliography}